\documentclass[twocolumn]{aa}
\usepackage{graphicx}
\usepackage{txfonts}

\begin{document}

   \title{Formation and evolution of dwarf elliptical galaxies}

   \subtitle{I. Structural and kinematical properties}

\author{S. De Rijcke \inst{1}\fnmsep\thanks{Postdoctoral Fellow of the
Fund for Scientific Research - Flanders (Belgium)(F.W.O.)},
D. Michielsen \inst{1}, H. Dejonghe \inst{1}, W.~W. Zeilinger
\inst{2}, \and G.~K.~T. Hau \inst{3} }

\offprints{S. De Rijcke}

\institute{Sterrenkundig Observatorium, Ghent University, Krijgslaan
281, S9, B-9000 Gent, Belgium, \\ \email{sven.derijcke@UGent.be} \and
Institut f\"ur Astronomie, Universit\"at Wien,
T\"urkenschanzstra{\ss}e 17, A-1180 Wien, Austria \and Department of
Physics, South Road, Durham, DH1 3LE, UK }

   \date{Received September 15, 1996; accepted March 16, 1997}

\abstract{This paper is the first in a series in which we present the
results of an ESO Large Program on the kinematics and internal
dynamics of dwarf elliptical galaxies (dEs). We obtained deep major
and minor axis spectra of 15 dEs and broad-band imaging of 22
dEs. Here, we investigate the relations between the parameters that
quantify the structure (B-band luminosity $L_B$, half-light radius
$R_{\rm e}$, and mean surface brightness within the half-light radius
$I_{\rm e} = L_B / 2 \pi R_{\rm e}^2$) and internal dynamics (velocity
dispersion $\sigma$) of dEs. We confront predictions of the currently
popular theories for dE formation and evolution with the observed
position of dEs in $\log L_B$ vs. $\log \sigma$, $\log L_B$ vs. $\log
R_{\rm e}$, $\log L_B$ vs. $\log I_{\rm e}$, and $\log R_{\rm e}$
vs. $\log I_{\rm e}$ diagrams and in the ($\log \sigma,\log R_{\rm
e},\log I_{\rm e}$) parameter space in which bright and
intermediate-luminosity elliptical galaxies and bulges of spirals
define a Fundamental Plane (FP). In order to achieve statistical
significance and to cover a parameter interval that is large enough
for reliable inferences to be made, we merge the data set presented in
this paper with two other recently published, equally large data sets.

We show that the dE sequences in the various univariate diagrams are
disjunct from those traced by bright and intermediate-luminosity
elliptical galaxies and bulges of spirals. It appears that
semi-analytical models (SAMs) that incorporate quiescent star
formation with an essentially $z$-independent star-formation
efficiency, combined with post-merger starbursts and the dynamical
response after supernova-driven gas-loss, are able to reproduce the
position of the dEs in the various univariate diagrams. SAMs with
star-formation efficiencies that rise as a function of redshift are
excluded since they leave the observed sequences traced by dEs
virtually unpopulated. dEs tend to lie above the FP and the FP
residual declines as a function of luminosity. Again, models that take
into account the response after supernova-driven mass-loss correctly
predict the position of dEs in the ($\log \sigma,\log R_{\rm e},\log
I_{\rm e}$) parameter space as well as the trend of the FP residual as
a function of luminosity.

While these findings are clearly a success for the
hierarchical-merging picture of galaxy formation, they do not
necessarily invalidate the alternative ``harassment'' scenario, which
posits that dEs stem from perturbed and stripped late-type disk
galaxies that entered clusters and groups of galaxies about 5~Gyr
ago.

}

\authorrunning{De Rijcke {\em et al.}}

   \maketitle
%

\section{Introduction}

It is known for almost two decades now that dynamically hot galaxies
(elliptical galaxies and bulges of spiral galaxies) are not scattered
randomly in the three-dimensional space spanned by B-band luminosity
($\log L_B$, { expressed in solar B-band luminosities}), half-light
radius ($\log R_{\rm e}$, { expressed in kiloparsecs}), and
velocity dispersion ($\log \sigma$, { expressed in km s$^{-1}$})
but that instead they occupy a slender plane:~the Fundamental Plane
(FP) (\cite{dd87}, \cite{dr87}, \cite{be92}). Projections of the FP
onto the coordinate planes, in combination with the particular way in
which galaxies are distributed within the FP, produce the univariate
relations between luminosity and velocity dispersion ($\log L_B = {\rm
const.} + 3.739 \log \sigma$, \cite{fj76}), luminosity and half-light
radius ($\log L_B = {\rm const.}  + 1.187 \log R_{\rm e}$,
\cite{fi63}), surface brightness { (expressed in solar B-band
luminosities per square parsec)} and half-light radius ($\log I_{\rm
e} = {\rm const.}  - 0.813 \log R_{\rm e}$, \cite{ko77}), and
luminosity and surface brightness ($\log L_B = {\rm const.}  -1.460
\log I_{\rm e}$, \cite{bi84}) (see \cite{gu93} for a compilation of
these so-called fundamental relations).

Dwarf elliptical galaxies (dEs) are small, low-luminosity galaxies
($M_B \gtrsim -18$~{ mag}) (\cite{fb}). They are among the most
numerous galaxy species in the universe and are found abundantly in
groups and clusters of galaxies. Their diffuse, { approximately}
exponentially declining surface-brightness profiles set them apart
from the compact ellipticals (cEs) which, while occupying the same
luminosity range as the dEs, have much higher { central} surface
brightnesses and a de Vaucouleurs-like surface-brightness profile. We
distinguish dwarf lenticular galaxies (dS0s) from dEs by their highly
flattened appearance (E6/E7) and their disky isophotes. The locus of
the dEs and dS0s in the ($\log \sigma,\log R_{\rm e},\log I_{\rm e}$)
space and in the univariate diagrams that require kinematical
information (such as the Faber-Jackson relation, hereafter FJR) was up
to now rather uncertain. After the early work by \cite{ni90},
\cite{be92}, and \cite{gu93}, relatively little attention has been
paid to the relations between the internal kinematics and the
structural parameters of these faint, small elliptical galaxies and,
particularly, to what we can learn from such relations regarding the
origin and evolution of dEs. This is in part due to their low surface
brightness, which makes spectroscopy, required for the extraction of
stellar kinematics, very time-consuming. This also limits the size of
the individual data sets and consequently compromises the statistical
significance of the results. Still, explaining the scaling relations
among the structural parameters of dwarf galaxies, which are believed
to be the building blocks of more massive galaxies, is a crucial test
for theories of cosmological structure formation. A comparison of
theoretical predictions with observations can put strong constraints
on the cosmological star-formation history and help to refine
prescriptions for e.g. star formation and energy feedback from
supernova explosions (which play a crucial role in { low-mass}
dwarf galaxies).

{ This paper is the first in a series in which we present the
results based on the kinematics of a sample of 15 dEs and dS0s and
photometry of 22 dEs/dS0s, both in group and cluster environments,
observed in the course of an ESO Large Program.} We explore the
relations between the various structural parameters of dEs and compare
them with theoretical predictions. This paper is organised as
follows:~we briefly discuss the observations and data reduction
procedures in section \ref{obs}. In section \ref{sectheory}, the
theoretical models with which we compare our data are presented. The
univariate relations between the different structural parameters are
discussed in section \ref{mono}. The position of the dEs with respect
to the Fundamental Plane is presented in section \ref{bivar}. We end
with a discussion of the results in section \ref{discuss}.


\section{Observations and data reduction} \label{obs}

Within the framework of an ESO Large Program, { we observed deep
major and minor axis spectra with unprecedented spatial and spectral
resolution of a sample of 15 dEs and dS0s, both in group (NGC5044,
NGC5898, and NGC3258 groups) and cluster environments (Fornax
cluster). We also collected Bessel VRI-band images of 22 dEs and dS0s
(including the 15 dEs/dS0s of the spectroscopic sample).} The targets
were selected because they were faint galaxies with elliptical
isophotes { (with absolute magnitudes in the range $-13.3 \ge M_B \ge
-18$~mag in order to distinguish them from the brighter ellipticals
and the fainter dwarf spheroidals (dSphs))} and had a high enough
surface brightness ($\mu_{\rm e,B} \le 23.5$~mag, with $\mu_{\rm e,B}$
the mean B-band surface brightness within the half-light radius) to
make the extraction of reliable kinematics out to roughly
$1.5-2\,R_{\rm e}$ feasible. In order to obtain a good spatial
sampling of the kinematics, the seeing FWHM should be much smaller
than a galaxy's half-light radius so galaxies with a half-light radius
smaller than $R_{\rm e} \approx 4''$ were rejected as possible
targets. There was no direct kinematical selection criterion.

The data were obtained with the FORS2 imaging spectrograph mounted on
the unit telescopes Kueyen and Yepun of the VLT. The images were
bias-subtracted and flatfielded using skyflats taken during twilight
of the same night as the science frames. The sky background was
removed by fitting a tilted plane to regions of the images free of
stars or other objects and subtracting it. The photometric zeropoints
in each band were measured using photometric standard stars observed
during the same night as the science frames. The images were corrected
for airmass and for interstellar extinction, using the Galactic
extinction estimates from \cite{schlegel98}. We measured the
surface-brightness profile, position angle, and ellipticity as a
function of the geometric mean of major and minor axis distance,
denoted by $a$ and $b$ respectively, using our own
software. Basically, the code fits an ellipse through a set of
positions where a given surface brightness level is reached (cosmics,
hot pixels, and foreground stars were masked and not used in the
fit). The shape of an isophote, relative to the best fitting ellipse,
is quantified by expanding the intensity variation along this ellipse
in a fourth order Fourier series with coefficients $S_4$, $S_3$, $C_4$
and $C_3$~:
\begin{eqnarray}
I(a,\theta) &=& I_0(a) \left[ 1 + C_3(a) \cos(3\theta)+ C_4(a)
\cos(4\theta)) \right.+ \nonumber \\ && \left. S_3(a)\sin(3\theta))+
S_4(a) \sin(4\theta) \right].
\end{eqnarray}
Here, $I_0(a)$ is the average intensity of an isophote with semi-major
axis $a$ and the angle $\theta$ is measured from the major axis. The
basic photometric parameters of the galaxies are presented in Table
\ref{tab1}. This smooth representation of a galaxy's
surface-brightness profile, $I(a,\theta)$, was integrated over
circular apertures { out to the last isophote we could reliably
measure in order to calculate its total magnitude. We did not wish to
extrapolate the surface brightness profile or the growth curve beyond
the region covered by the data (e.g. by using a S\'ersic profile)
because of the uncertainties inherent to this procedure. However, our
images go down at least 8~mags. Cutting off an exponential surface
brightness profiles 8~mags below its central value results in an
underestimate of the total luminosity by less than one percent (a
cut-off at 6~mags below the central surface brightness, as in a few
less deep images, results in an underestimate of the total luminosity
by about 2.5\% for an exponential profile).}

\begin{table*}
\caption{Structural properties of the observed dEs.The FCC dEs are
members of the Fornax Cluster (\cite{fe89}); the FS dEs belong to the
NGC5044 and NGC3258 groups and are cataloged by \cite{fe90}. {
NGC5898\_DW1 and NGC5898\_DW2 are two previously uncataloged dEs in
the NGC5898 group (where DW stands for DWARF).} For each galaxy, the
equatorial coordinates, the B-band apparent magnitude $m_B$ (taken
from \cite{fe89} and \cite{fe90}, except for FCC046 and FCC207 for
which we have obtained B-band images), { the luminosity-weighted
ellipticity $\epsilon = 1- q$, with $q$ the isophotal axis ratio,
outside the inner 1~arcsec where the ellipticity measurement is
affected by seeing}, the { luminosity-weighted} velocity dispersion
$\sigma$~(km~s$^{-1}$, { see eq. (\ref{sigmeq2}) for a precise
definition of $\sigma$}), the half-light radius $R_{\rm e}$~(arcsec,
{ this is the radius of the circular aperture that encloses half of
the light}), derived from the R-band images, and the B-band mean
surface brightness within one half-light radius $\mu_{{\rm e},B}$ are
listed. { We used our R-band $R_{\rm e}$ to calculate $\mu_{{\rm
e},B}$. This does not influence our results in any way since we found
$R_{\rm e}$ not to differ between R, B, and V-band images.} For 7 dEs,
only photometry is available.
\label{tab1}}
\begin{center}
\begin{tabular}{|l|cc|c|c|c|c|c|} \hline
  name   & $\alpha$ (J2000)   & $\delta$ (J2000) & $m_B$    &  $\epsilon$          & $\sigma$      & $R_{\rm e}$   & $\mu_{{\rm e},B}$ \\ \cline{1-8}
  FCC043 & 03 26 02.2         &  -32 53 40       & 13.91    &   0.26               & 56.4$\pm$3.7  &	16.9       &  22.05       \\
  FCC046 & 03 26 25.0         &  -37 07 41       & 15.99    &   0.36               & 61.4$\pm$5.0  &	6.7        &  22.12\\
  FCC100 & 03 31 47.6         &  -35 03 07       & 15.30    &   0.28               &    /          &    11.7       &  22.63 \\
  FCC136 & 03 34 29.5         &  -35 32 47       & 14.81    &	0.21               & 64.3$\pm$3.8  &	14.2       &  22.57  \\
  FCC150 & 03 35 24.1         &  -36 21 50       & 15.70    & 	0.19               & 63.8$\pm$3.9  &	5.7        &  21.48 \\
  FCC188 & 03 37 04.5         &  -35 35 26       & 16.10    &	0.06               &  /		   &    9.3        &  22.94 \\
  FCC204 & 03 38 13.6         &  -33 07 38       & 14.76    &	0.61               & 67.2$\pm$4.4  &	11.5       &  22.06   \\
  FCC207 & 03 38 19.3         &  -35 07 45       & 16.19    &	0.15               & 60.9$\pm$6.6  & 	8.4        &  22.81 \\
  FCC245 & 03 40 33.9         &  -35 01 23       & 16.00    & 	0.11               & 39.5$\pm$4.2  &	11.4       &  23.28  \\
  FCC252 & 03 40 50.4         &  -35 44 54       & 16.00    &	0.08               & /		   &    7.9        &  22.47\\
  FCC266 & 03 41 41.4         &  -35 10 12       & 15.90    &	0.11               & 42.4$\pm$3.4  & 	7.1	   &  22.15  \\
  FCC288 & 03 43 22.6         &  -33 56 25       & 15.10    &	0.72               & 48.5$\pm$3.3  &	9.5        &  21.99  \\
  FCC303 & 03 45 14.1         &  -36 56 12       & 15.50    &   0.09               & /             &    11.6       &  22.81 \\
  FCC316 & 03 47 01.4         &  -36 26 15       & 16.30    &   0.26               & /             &    9.0        &  23.06  \\
  FCC318 & 03 47 08.2         &  -36 19 36       & 16.10    &	0.18               & /		   &    10.6       &  23.22 \\
  FS029  & 13 13 56.2         &  -16 16 24       & 15.70    &	0.54               & 59.6$\pm$3.6  & 	8.9        &  22.44 \\
  FS075  & 13 15 04.1         &  -16 23 40       & 16.87    &   0.10               & /             &    6.8        &  23.03\\
  FS076  & 13 15 05.9         &  -16 20 51       & 16.10    &   0.07               & 56.8$\pm$3.8  & 	4.4        &  21.41 \\
  FS131  & 13 16 49.0         &  -16 19 42       & 15.30    &	0.54               & 87.0$\pm$3.2  & 	8.1        &  21.83 \\
  FS373  & 10 37 22.9         &  -35 21 37       & 15.60    & 	0.23               & 73.1$\pm$3.1  & 	7.9        &  22.03 \\
  NGC5898\_DW1    & 15 18 13.0         &  -24 11 47       & 15.66    &	0.34               & 43.5$\pm$3.0  &	8.7        &  22.35 \\
  NGC5898\_DW2    & 15 18 44.7         &  -24 10 51       & 16.10    & 	0.57               & 44.2$\pm$3.4  &	5.9        &  21.95  \\ \cline{1-8}
\end{tabular}
\end{center}
\end{table*}
The spectra, with typical exposure times of $5-8$~h per position angle
and a seeing in the range $0.3''-1.0''$~FWHM, cover the wavelength
region around the strong Ca{\sc ii} triplet absorption lines ($\sim
8600$~{\AA}). All standard data reduction procedures
(bias-subtraction, flatfielding, cosmic removal,
wavelength-calibration, sky-subtraction, flux-calibration) were
carried out with {\tt ESO-MIDAS}\footnote{{\tt ESO-MIDAS} is developed
and maintained by the European Southern Observatory}, {\tt
IRAF}\footnote{{\tt IRAF} is distributed by the National Optical
Astronomy Observatories, which are operated by the Association of
Universities for Research in Astronomy, Inc., under cooperative
agreement with the National Science Foundation.}, and our own
software. We fitted the dispersion relation with a cubic spline, which
rectified the lines of the arc spectra to an accuracy of $1-2$~km~s$^{-1}$
FWHM. We extracted the stellar kinematical information by fitting a
weighted mix of late G to late K giant stars, broadened with a
parameterised line-of-sight velocity distribution (LOSVD) to the
galaxy spectra. We approximated the LOSVD by a fourth-order
Gauss-Hermite series (\cite{gh,vf}) (the kinematics of the full sample
and a complete discussion of the data acquisition and analysis will be
presented in the next paper in this series, De Rijcke {\em et al.}, in
prep.). The strong Ca{\sc ii} lines, which contain most of the
kinematical information, are rather insensitive to the age and
metallicity of an old stellar population (see e.g. \cite{mi03},
\cite{fem03}, \cite{sa02}), so template mismatch does not
significantly affect our results. The spectra contain useful
kinematical information out to 1.5-2 $R_{\rm e}$.


\section{Theoretical models for dE formation} \label{sectheory}

{ Here, we discuss the models for dE formation and evolution which
in the following sections will be compared with the data. First, three
different calculations are introduced that are all based on the idea
that dEs are primordial objects that lost their gas in a
supernova-driven galactic wind (a.k.a. the ``wind model'') and second,
we discuss model predictions based on the harassment scenario.}

\subsection{The ``wind model''}

The first detailed models of the evolution of galaxies taking into
account the dynamical response after a supernova-driven galactic wind
(hence the name of this class of models) were presented by \cite{ya87}
(hereafter YA87). These models follow the evolution of stellar systems
using one-zone models without internal structure and with masses
between $10^{5}\,M_\odot$ and $2 \times 10^{12}\,M_\odot$. They assume
an empirical scaling law between the gravitational binding energy
$\Omega$ and total mass $M$:~$\Omega \propto M^{1.45}$ (\cite{fi64},
\cite{sa79}). This translates into a mass-gravitational radius
relation of the form $R_G \propto M^2/\Omega \propto M^{0.55}$, a
mass-density relation $\rho \propto M/R^3 \propto M^{-0.65}$, and a
velocity dispersion-mass relation $\sigma \propto \sqrt{\Omega/M}
\propto M^{0.225}$ {\em before} any mass-loss has occurred. The
star-formation timescale $t_{\rm SF}$ is proportional to the minimum
of the collision time of molecular clouds and the free-fall time of
individual clouds. In both cases, $t_{\rm SF} \propto 1/\sqrt{\rho}$
with the proportionality constant chosen such that the present-day
color-magnitude relation, from giant ellipticals down to globular
clusters, is reproduced. The structure of these models is altered as
they expand in response to the loss of a mass-fraction $f$ in the form
of supernova-driven galactic winds. The value of $f$ is determined by
the timing of the galactic winds, with later winds expelling a smaller
gas fraction than earlier ones. In these models, the time-scales on
which all physical processes operate are determined by $M$, making $f$
also a function of $M$. The crux is, of course, for $f$ to have the
correct behaviour as a function of $M$ in order to reproduce the
observed relations between the different structural parameters of
spheroidal galaxies. Since $R_G$ is roughly proportional to $R_{\rm
e}$ in a constant-$M/L$ model, and the precise value of the radius
does not affect the physics of the models anyway, we scaled the
mass-radius relation such that the present-day half-light radii of
ellipticals with a luminosity $\log L_B = 10.4$ ($M_B = -20.5$~mag)
are reproduced. This is the boundary between, roughly speaking, the
disky, { cuspy}, isotropic intermediate-brightness ellipticals and
the boxy, anisotropic bright ellipticals { with a core}
(\cite{be89}, {\cite{gr03b}}).

More recent semi-analytical models (SAMs) (\cite{so01}, \cite{ny04})
take into account the hierarchical merger tree that leads up to the
formation of a galaxy of a given mass in a $\Lambda$CDM cosmology
($\Omega_{\rm baryon} \approx 0.02$, $\Omega_{\rm matter}=0.3$,
$\Omega_\Lambda=0.7$, $H_0=70$km~s$^{-1}$~Mpc$^{-1}$). SAMs make use
of prescriptions for star-formation, energy feedback from supernova
explosions, gas cooling, tidal stripping, dust extinction, and the
dynamical response to starburst-induced gas ejection and, despite the
inevitable oversimplifications in the description of immensely complex
processes such as star formation, they are able to account pretty well
for many observed properties of galaxies. E.g., \cite{so01} compare
their SAMs with the comoving number density of galaxies, the evolution
of the galaxy luminosity function and the H{\sc i} content of the
universe as a function of redshift, and the star-formation history of
the universe (the ``Madau diagram''). Based on this comparison, these
authors favor the idea that merger-induced star bursts have played an
important role in enhancing the star-formation efficiency at high
redshift. \cite{ny04} (NY04) moreover take into account the gas
ejection and the ensuing dynamical response of a galaxy after a
merger-induced star-burst. These authors compare the galaxies formed
in their SAMs with the observed properties of dynamically hot galaxies
(from the faint Local Group dSphs up to the very brightest elliptical
galaxies) and come to roughly the same conclusions as
\cite{so01}. Despite their simplicity, the YA87 models already capture
a lot of the physics of galaxy formation, at least in the mass-regime
of the dEs, which { apparently} formed while their progenitors were
still largely gaseous. {\em In all figures, we will use them as an
instructive proxy for the physically more motivated SAMs of
e.g. \cite{so01} and \cite{ny04}.} Using the tables given in
\cite{ya87}, we fitted cubic splines to all relevant parameters as
functions of $M$ in order to be able to plot them as continuous
sequences in the figures in the following sections.

In \cite{cc02} (hereafter CC02), the evolution of spherical elliptical
galaxies in a standard CDM universe ($\Omega_{\rm baryon}=0.1$,
$\Omega_{\rm dark}=0.9$, $\Omega_\Lambda=0$) is followed using
N-body/SPH simulations with prescriptions for star-formation,
feedback, cooling, and chemical evolution. A protogalaxy is assumed to
virialize at a redshift $z=5$, when it had an overdensity $1+\delta
\sim 200$, after which star-formation ensues. Two series of models,
with masses between $10^{8}$ and $10^{13}\,M_\odot$ and with different
initial densities, are presented. Models of series A have a high
initial density (they virialized at redshift $z=5$, when the universe
was still very dense), models of series B have low initial densities
(they virialized at much lower redshift). These models allow for a
more realistic description of the galactic wind as compared to
YA87:~only gas particles with velocities above the local escape speed
are blown away. This reduces the amount of mass that is lost and leads
to a much milder structural evolution compared to the YA87
models. Contrary to the YA87 models, dwarf galaxies are found to have
the longest star formation histories. Massive ellipticals hold on very
strongly to their gas reservoirs, allowing the gas to be converted
almost completely into stars in a single burst. In dwarf galaxies, on
the other hand, supernova explosions disperse the gas, switching off
further star formation. Gas cools, sinks back in and the
star-formation efficiency goes up again after which subsequent
supernova explosions disperse the gas and so on. This gives rise to
very inefficient, oscillating but long-lived star formation. { As
aforementioned, the way star formation is implemented in the YA87
models causes the star-formation efficiency to be highest in low-mass
galaxies, they simply do not continue forming stars for very long. By
the time a galactic wind sets off, most of the gas has been converted
into stars. Very massive galaxies form stars most slowly but they hang
onto their gas very strongly, leaving enough time for most of the gas
to be converted into stars. Hence, galaxies around $\log(L_B) = 7$
($M_B=-12$~mag) experience the most drastic mass-loss while more and
less luminous galaxies are less affected by gas ejection. This effect
is most likely due to the extreme simplicity of the models. The more
complex models, with more realistic star-formation histories, do not
show this behaviour. 

Both the YA87 and CC02 models do not take into account mergers. As a
consequence, they will underestimate the velocity dispersions and
half-light radii and overestimate the surface brightnesses of massive
ellipticals (see e.g. \cite{he92} and \cite{da03} for the effects of
dissipationless mergers and \cite{he93} and \cite{be98} for merger
simulations taking into account the presence of gas). SAMs (e.g. NY04)
follow the merger tree leading up to a galaxy (including
dissipationless mergers) and hence do a much better job at reproducing
the characteristics of massive galaxies. Hence, the failure of the
YA87 and CC02 models for bright galaxies does not signal a problem
with CDM. These models simply lack the physics necessary to explain
the properties of bright ellipticals.}

The N-body/SPH models presented by \cite{ka99} and \cite{ka01} (K01)
are based roughly on the same premises as those of CC02, although the
details differ. Using an N-body/SPH code, the evolution of a slowly
rotating (spin parameter $\lambda=0.02$) region in a standard CDM
universe, on which small-scale density fluctuations are laid out, is
followed from $z=40$ up to the present. Overdense clumps within this
region collapse, merge, and finally form a single elliptical
galaxy. Regions with masses between $10^{11}$ and $8 \times
10^{12}\,M_\odot$ are modeled. E.g., for a region with a total mass
$M=8 \times 10^{11}\,M_\odot$, star formation starts in merging clumps
at $z \approx 3.5$ and the final galaxy is in place at $z \approx
1.9$. No merging takes place after $z \approx 0.8$. The influence of
the energy feedback efficiency of supernova explosions (SN{\sc ii} and
SN{\sc i}a) on the evolution of elliptical galaxies and bright dEs is
explored. The more efficient the supernovae inject kinetic energy into
the interstellar medium, the sooner the gas is blown away. This
results in more expelled gas and more structural evolution in low-mass
galaxies. Two series of models are calculated:~a series with very
efficient kinetic feedback, which is able to reproduce the observed
color-magnitude relation of elliptical galaxies, and one with minimal
feedback. As an example: a bright dwarf galaxy with an initial total
mass $M=10^{11}\,M_\odot$ and with a high feedback efficiency develops
a galactic wind at $z\approx 1.7$. By $z \approx 1.2$, most of the gas
has been blown out. For these models, only photometric quantities are
calculated.

{ It should be noted that CC02 and K01 use a CDM cosmology whereas
NY04 use $\Lambda$CDM. This is probably not a major concern. CC02
follow the evolution of an isolated galaxy, disconnected from the
universal expansion. The only cosmological parameter that goes into
these calculations is the density of the universe at the virialization
redshift of the galaxy. However, at $z \sim 5$, a CDM universe very
much resembles a $\Lambda$CDM universe since $\Omega_\Lambda$ started
to dominate $\Omega_{\rm matter}$ only at $z \sim 0.3$. In K01, most
of the merging activity, which is influenced by the dynamics of the
universe, has ended around $z=2$. So, hierarchical merging takes place
at a cosmic time when CDM and $\Lambda$CDM universes do not differ
much. The YA87 models, due to their simplicity, are essentially devoid
of cosmological considerations. Moreover, slight effects due to
cosmology can probably be offset by using slightly different
star-formation or feedback efficiencies. This makes us confident that
the different cosmologies do not yield significantly different
predictions for the properties of individual galaxies.}

\subsection{The harassment scenario}

\begin{figure*}
\vspace*{8.4cm} \special{hscale=52 vscale=52 hsize=700 vsize=245
hoffset=-42 voffset=-140 angle=0 psfile="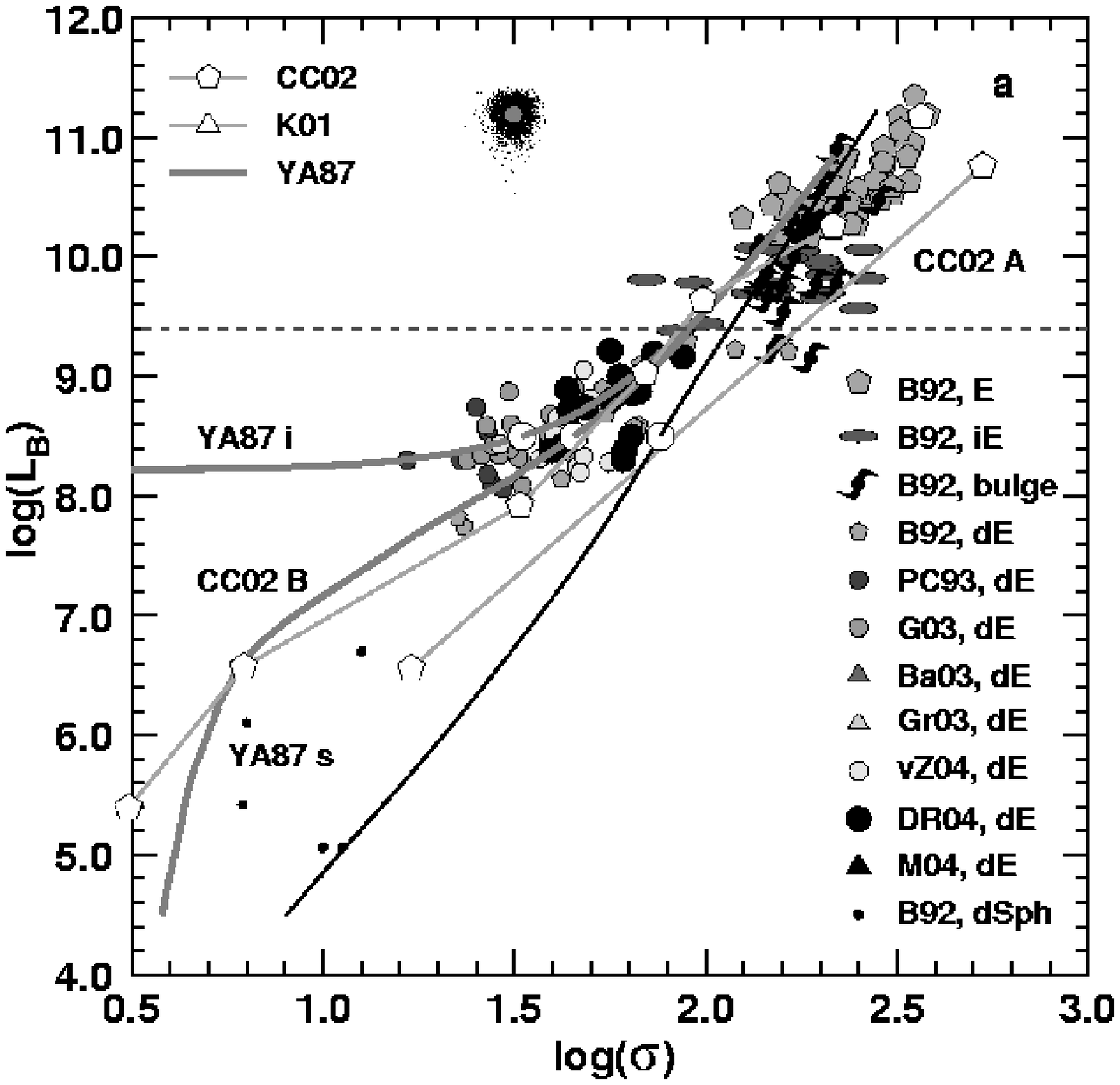"}

\special{hscale=52 vscale=52 hsize=550 vsize=245 hoffset=235
voffset=-140 angle=0 psfile="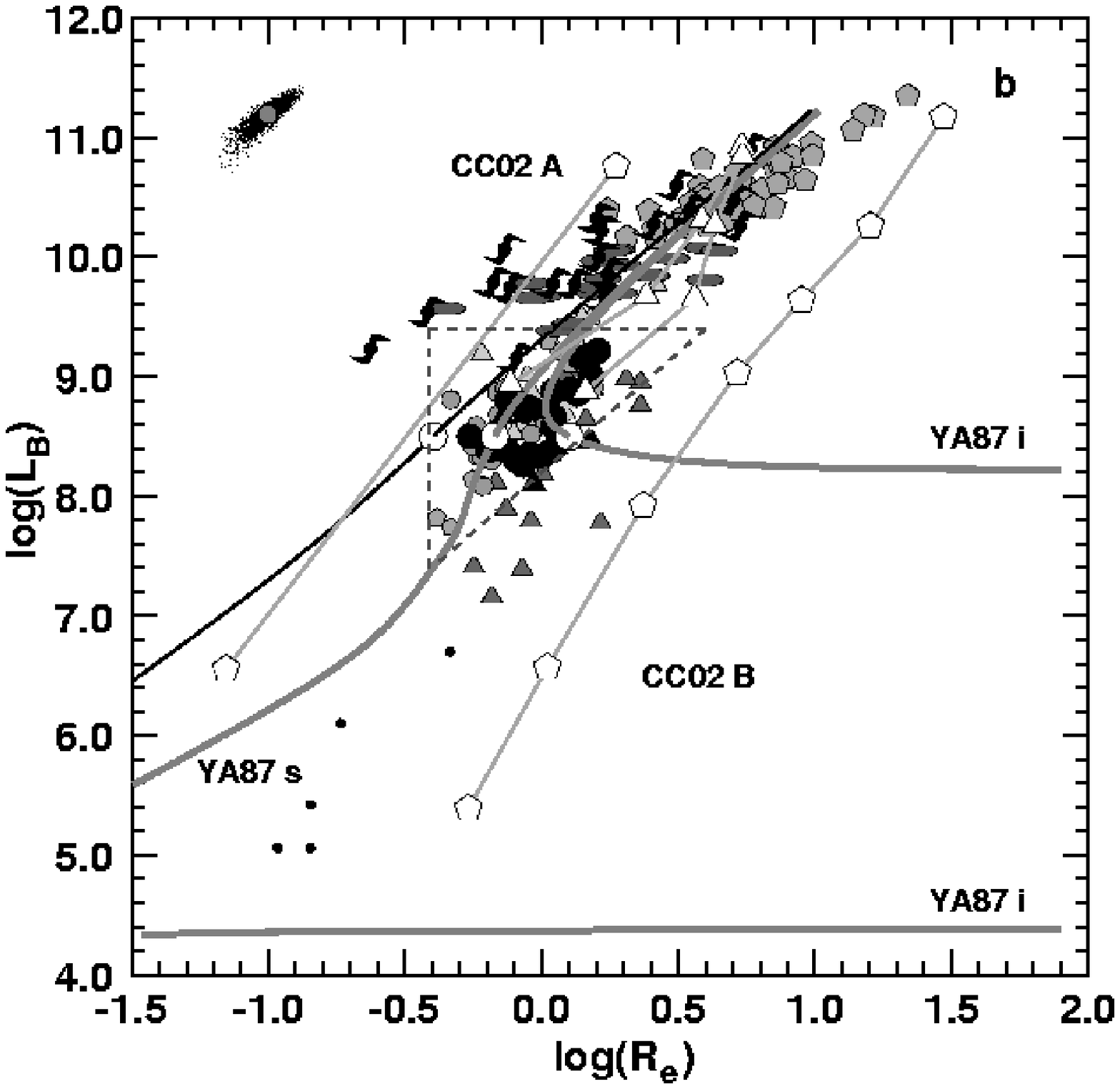"}

\vspace*{8.3cm} \special{hscale=52 vscale=52 hsize=900 vsize=250
hoffset=-42 voffset=-150 angle=0 psfile="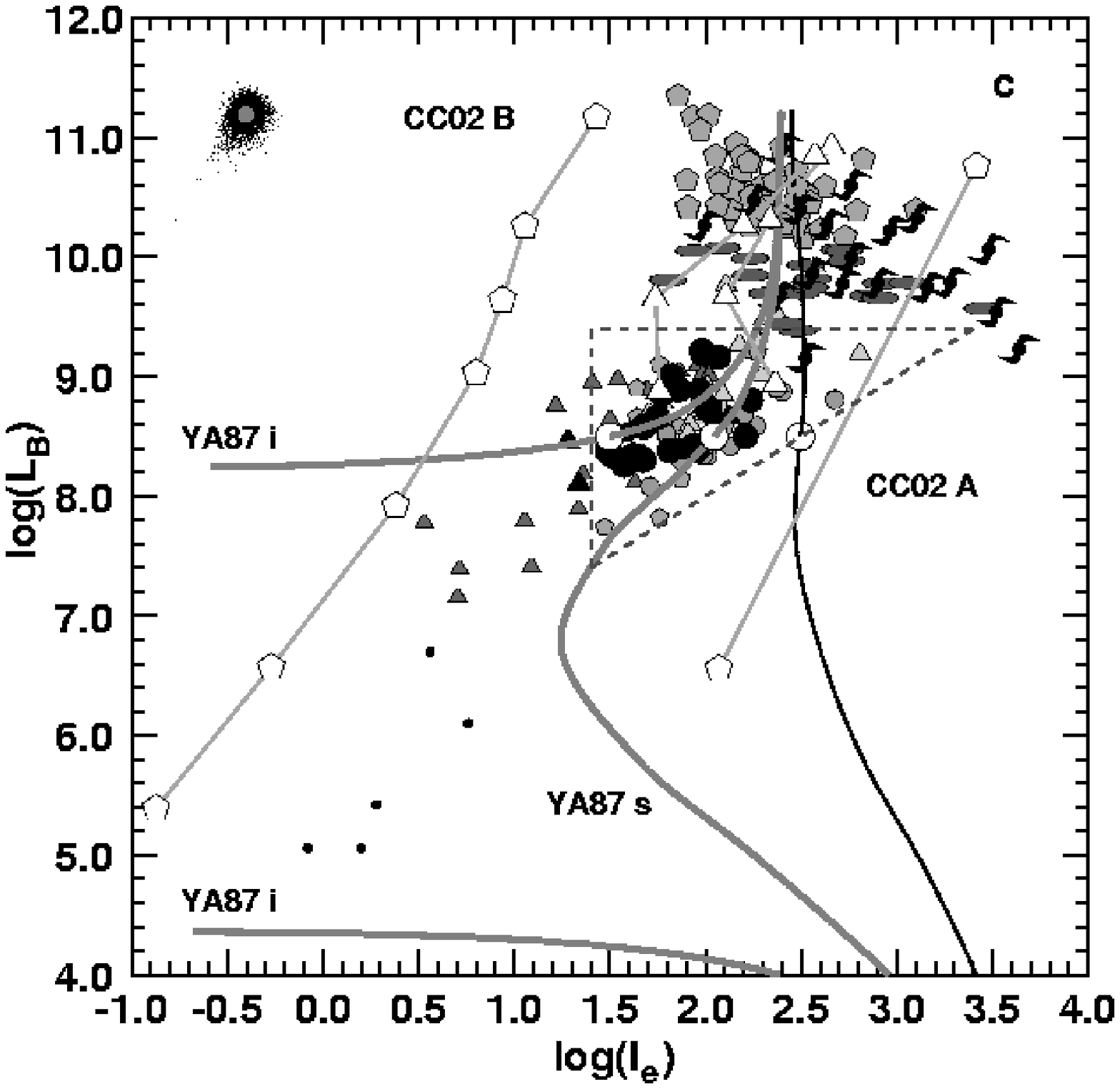"}

\special{hscale=52 vscale=52 hsize=550 vsize=250 hoffset=235
voffset=-150 angle=0 psfile="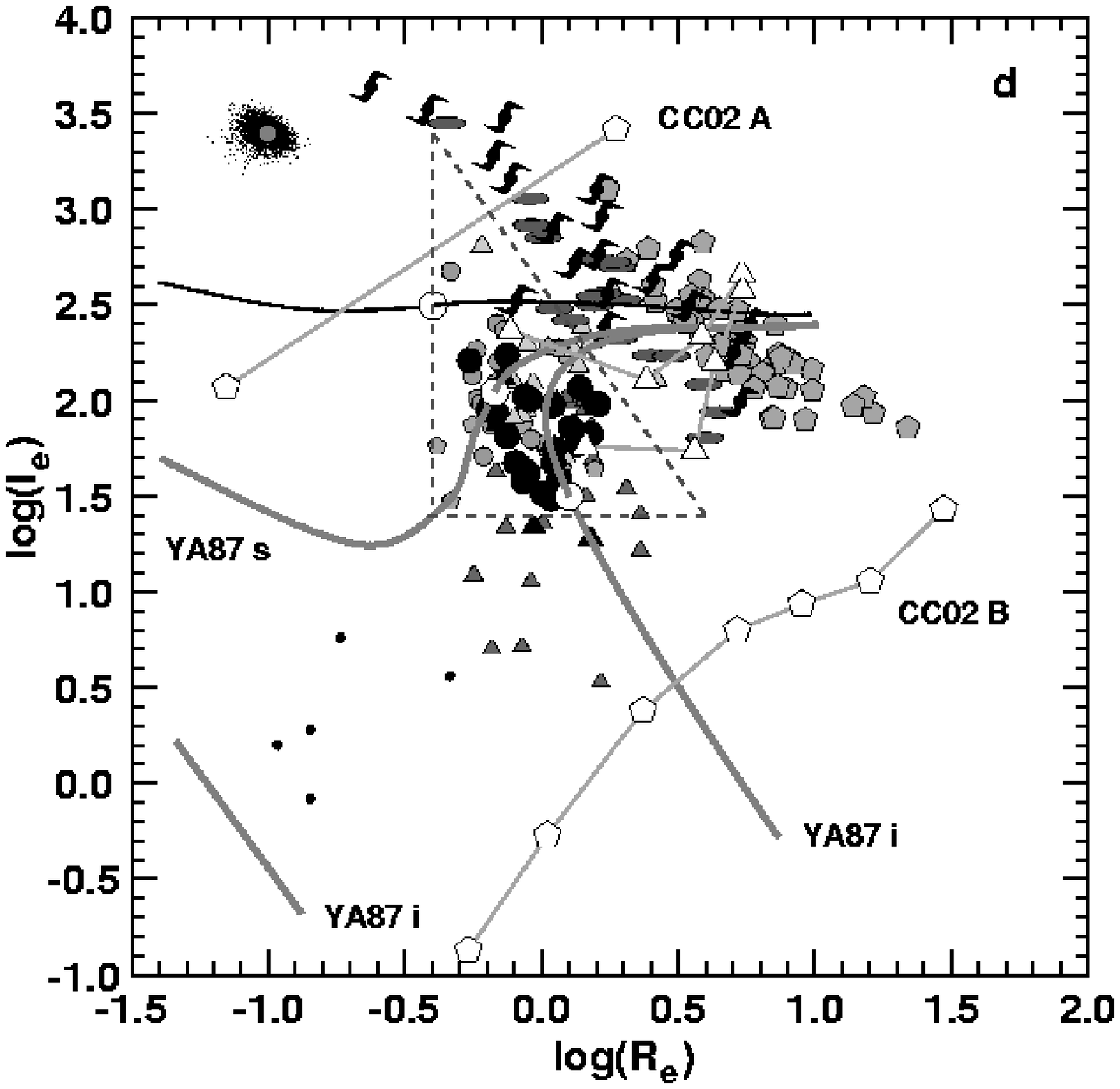"}

\caption{Panel {\bf a}:~the $\sigma-L_{\rm B}$ relation; panel {\bf
b}:~the $R_{\rm e}-L_{\rm B}$ relation; panel {\bf c}:~the $I_{\rm
e}-L_{\rm B}$ relation; panel {\bf d}:~the $R_{\rm e}-I_{\rm e}$
relation. The regions in these diagrams that satisfy our selection
criteria 
are delineated by dotted lines. { $L_B$ is expressed in solar
B-band luminosities, $\sigma$ in km~s$^{-1}$, $R_{\rm e}$ in
kiloparsecs, and $I_{\rm e}$ in solar B-band luminosities per square
parsec.} The symbols representing the various datasets, galaxy types
(bright ellipticals:~E, intermediate-luminosity ellipticals:~iE,
bulges of spiral galaxies:~bulge, dwarf ellipticals:~dE, dwarf
spheroidals:~dSph), and theoretical models are indicated in panel {\bf
a}. The black line traces the zero mass-loss relation of the YA87
models. The thick grey lines trace the relations after slow (YA87 s)
or instantaneous (YA87 i) gas removal. The position of a typical dE
with apparent magnitude $m_B = 16$ at the distance of Fornax before
and after supernova-driven mass-loss is indicated with white
circles. 
The CC02 models (white pentagons) of series A (CC02 A) collapse to a
final state with a very small half-light radius and a large velocity
dispersion. The models of series B (CC02 B) on the other hand
reproduce the observed $\sigma-L_{\rm B}$ relation but have half-light
radii that are too large. On the whole, the CC02 models are able to
reproduce the observed slopes of the several structural relations. The
photometric properties of the K01 models (white triangles) {
roughly agree with the observations and the YA87 models}. For a given
mass, the trend for K01 models as a function of feedback efficiency
parallels that of the YA87 models as a function of mass-loss
timescale. In the upper left corner of each panel, a cloud of 3000
simulated { data points} gives an idea of the systematic and
statistical uncertainty on a typical { DR04} data point (indicated
by a grey dot).
\label{LBsigma} }
\end{figure*}

Alternatively, dEs could stem from late-type disk galaxies that
entered the clusters and groups of galaxies about 5~Gyr ago { ($z
\sim 0.5$)}. $N$-body simulations show that gravitational interactions
trigger bar-formation in any small late-type disk galaxy (Scd-Irr)
orbiting in a cluster (\cite{mo96}, \cite{mo98}) or around a massive
galaxy in a group environment (\cite{ma01}) and strip large amounts of
stars, gas, and dark matter from it by tidal forces. Internal
dynamical processes (such as the buckling of the bar) subsequently
transform a disk galaxy into a dynamically hot spheroidal dE within a
timespan of a few Gyr { (\cite{ma04})}. Gas (if it is not stripped
off by the ram pressure of the intergalactic medium, see
e.g. \cite{ma03}) is funneled inwards by the non-axisymmetric force
field of the bar { (\cite{ma01})}. If this gas is converted into
stars, the resulting galaxy is likely to have a higher central surface
brightness than its progenitor { and would probably resemble a
nucleated dE. Purely stellar dynamical processes such as bar formation
and buckling lead to a more shallow central density enhancement}. The
dEs formed in these simulations rotate quite rapidly and some still
display some memory of their former state. Examples are dEs with
embedded stellar disks (\cite{ba02,dr03a,gr03}) or kinematically
decoupled cores (\cite{dr04}). This model explicitly takes into
account the fact that dEs are found predominantly in clusters and
groups of galaxies. It also offers a natural explanation for the
Butcher-Oemler effect (\cite{bo78}):~at $z \sim 0.4$, a population of
blue, distorted galaxies is present in clusters of galaxies while no
such objects reside in present-day clusters. The former could be
late-type galaxies caught in the predicament of being harassed and
transformed into more spheroidal objects.

Clearly, the parameter space that should be explored to make firm
predictions based on this scenario is huge. One should start with
gas-rich late-type galaxies of various masses and with different
scale-lengths, placed on different orbits in a variety of cluster and
group environments, using realistic prescriptions for star-formation,
energy feedback, gas cooling, and ram-pressure stripping by the
intra-cluster medium. As long as this formidable task has not been
performed, we will have to settle with what can be gleaned from the
few calculations that have been published up to now. These show that,
harassment being a stochastic process whose effects depend in a very
complicated way on the orbit of a progenitor galaxy through a cluster,
it is rather unlikely that dEs end up close to a slender plane like
the FP. The half-light radius of a dE formed through harassment is
about half that of its progenitor, with the central surface brightness
about two to three times higher than the original value
(\cite{ma01}). The velocity dispersion has also roughly doubled. {
Moreover, it is very likely that interactions induce star-bursts and
hence that dEs formed through harassment harbor stellar populations of
different ages, further adding to the expected scatter about the FP
(\cite{fo89}).}  Although more simulations are needed to converge on a
coherent picture of the effects of harassment, two predictions can
already be made:~(1) some dEs should still contain some memory of
their disky, gaseous past and (2) only weak correlations with large
scatter are expected between structural parameters. The latter is due
to the stochastic nature of harassment and, if the Butcher-Oemler
galaxies at $z\sim 0.4$ are anything like the present-day Scd-Irr
galaxies, due to the large scatter of the progenitor galaxies in the
FP (\cite{bu97}).

{ Yet another possible way to produce dwarf galaxies is described
by \cite{pk98} or \cite{du04}. The hundreds of young compact massive
star clusters formed during the merger of two gaseous disk galaxies
may coalesce within a few 100~Myr to form a small number of objects,
with masses of order $10^9\,M_\odot$, with negligible dark-matter
content, and with a half-mass radius of a few 100~pcs. However,
simulations have thus far not been able to make predictions for the
present-day properties of these so-called tidal dwarf galaxies. Also,
major mergers cannot have produced a significant fraction of the dE
population since each merger only yields one or two tidal dwarf
galaxies. For these reasons, and since the objects formed this way are
less massive than the dEs considered here and in many respects are
more like the Local Group dSphs, we will not discuss this formation
mechanism here.}


\section{Univariate relations} \label{mono}

In the following, we confront the predictions of the models discussed
in section \ref{sectheory}, with the observed relations between the
structural and kinematical parameters of dynamically hot galaxies. The
most massive representatives of this family of galaxies are the giant
ellipticals (B-band luminosity $\log L_B \sim 11.5$ { ($M_B =
-23.3$~mag)}); the least massive systems are the Local Group dwarf
spheroidals ($\log L_B \sim 6$ { ($M_B = -9.5$~mag)}). Three data
sets of kinematics of dEs in the luminosity range $\log L_B \approx 8
- 9.2$ { ($M_B = -14.5$ to $-17.4$)} are currently available:~our
sample of 15 group and cluster dEs/dS0s (DR04), the sample of 17 Virgo
cluster dEs observed by \cite{ge03} (G03) and the 16 Virgo cluster dEs
from van Zee {\em et al.}, 2004 (vZ04) (which has an overlap of 5
objects with the G03 sample). Since none of the three datasets on its
own can boast statistical significance or cover a parameter interval
large enough for reliable inferences to be made, it is natural to
combine them, bringing the total number of dEs with spatially resolved
kinematics to 43. For completeness, we also compare our results with
the central velocity dispersions of nucleated dEs measured by
\cite{pc93} (PC93).

In diagrams relating only photometrical parameters, we have included
the sample of 25 Virgo cluster dEs and dS0s of \cite{ba03} (Ba03),
which was used for the determination of surface-brightness fluctuation
(SBF) distances, the Fornax dEs presented in \cite{mi04} (M04){,
and the 18 Coma dEs analysed in \cite{gr03b} (Gr03).} The dwarf
galaxies in the Ba03 sample were selected because they have large
half-light radii and yield as many independent SBF measurements as
possible. Hence, the Ba03 sample traces the dE sequences in the
photometric univariate diagrams in regions as yet unexplored by
spectroscopic samples. The relevant data for giant ellipticals,
intermediate-luminosity ellipticals, bulges, dwarf spheroidals and a
handful of dEs come from \cite{be92} (B92). Where necessary, all
datasets have been converted to B-band luminosities assuming ${\rm
B}-{\rm V}=0.7$ and to a Hubble parameter $H_0 =
70$~km~s$^{-1}$~Mpc$^{-1}$ (\cite{fr01}). This places the Fornax
cluster at a distance of 19.7~Mpc ($v_{\rm sys} = 1379$~km~s$^{-1}$)
and the Virgo cluster at 15.4~Mpc ($v_{\rm sys} = 1079$~km~s$^{-1}$),
both in good agreement with SBF distances:~\cite{je03} and \cite{je04}
find a Fornax distance of $20.3 \pm 0.7$~Mpc and a distance of $15.8
\pm 1.4$~Mpc for the Virgo M87-subcluster, with the main uncertainty
being the depth of the Virgo cluster. Using this $H_0$ value, the
NGC5044 group distance is estimated at 35.1~Mpc ($v_{\rm sys} =
2459$~km~s$^{-1}$), that of NGC3258 at 40.7~Mpc ($v_{\rm sys} =
2848$~km~s$^{-1}$), and NGC5898 at 30.3~Mpc ($v_{\rm sys} =
2122$~km~s$^{-1}$). Using SBF, these groups or their dominant
elliptical galaxy are placed at $31.2 \pm 4.0$~Mpc (NGC5044), $32.1
\pm 4.0$~Mpc (NGC3258), and $29.1 \pm 3.5$~Mpc (NGC5898)
(\cite{to01}). Hence, with the exception of the NGC3258 group,
Hubble-distances and SBF-distances agree within the errorbars.

Below, we discuss the position of the dEs in the (mutually dependent)
$\sigma-L_{\rm B}$, $R_{\rm e}-L_{\rm B}$, $I_{\rm e}-L_{\rm B}$, and
$I_{\rm e}-R_{\rm e}$ diagrams. Since the quantities plotted on the
ordinate and abscissa are not necessarily independent (e.g. $I_{\rm
e}$ and $R_{\rm e}$), their errors can also be correlated. We
estimated the systematic and statistical { (measurement)}
uncertainties on all quantities for a typical galaxy in our sample by
assuming the values $\sigma = 50 \pm 3$~km~s$^{-1}$ (assumed to be a
1$\sigma$ Gaussian error), $R_{\rm e} = 1.25 \pm 0.06$~kpc, $L_B = (8
\pm 0.8) \times 10^8 \,L_{\odot,B}$. { Using a Gaussian
random-number generator, a data-point can be generated that is
afflicted only by measurement errors. For each such new data-point, a
distance estimate was generated, assuming a distance of $30 \pm 3$~Mpc
(1$\sigma$-error), which, even in the absence of other measurement
errors, would introduce a correlated uncertainty on distance-dependent
quantities. Each new data-point was then shifted according to its
distance. We calculated 10000 new data-points for this typical galaxy,
each one affected randomly by measurement errors and a
distance-dependent shift. This way, the Monte-Carlo procedure properly
adds the random measurement errors and the systematic
distance-dependent errors.} In the corner of each panel of
Fig. \ref{LBsigma}, a dark grey data point is indicated, surrounded by
3000 of these simulated data points (small black dots). These give an
idea of the typical systematic and statistical (correlated) errors on
our data. They over-estimate the true scatter between the individual
data points since all galaxies belonging to the same group or cluster
will be affected in the same way by the distance uncertainty.

\subsection{The $\sigma-L_{\rm B}$ or Faber-Jackson relation (FJR)}

Bright and intermediate-luminosity ellipticals and bulges of spiral
galaxies adhere closely to the FJR:~$L_B \propto \sigma^\alpha$, with
$\alpha \sim 4$ (\cite{fj76}). { \cite{hel92} fitted a straight line
throught data of 4 dEs (three Local Group dEs and one Virgo dE) and 4
Local Group dSphs and found $L_B \propto \sigma^{2.5}$. They however
omitted one dE from their original dataset of five and two dSphs from
an original sample of six. All three deviate significantly from the
relation fitted to the eight galaxies that were used, casting serious
doubts on the slope of the fitted relations.} PC93 fitted a straight
line to the then available dE and dSph data and found $L_V \propto
\sigma^{5.6 \pm 0.9}$. \cite{gu93} connected the B92 dE data with that
of the dSphs and found a similar slope to the FJR of the bright
ellipticals but with a different zero-point. In both attempts to
measure the dE FJR, the region of about 5 magnitudes in luminosity
between the dEs and the dSphs, for which no data was available, had to
be bridged.

We are now for the first time in the situation that kinematical data
of enough dEs are available to measure the dE FJR without the need of
extrapolating towards the dSphs. Many dEs contain a central brightness
peak, called the nucleus. Hence, the central velocity dispersion does
not necessarily reflect the dynamics of the whole galaxy and cannot be
expected to be a good measure for a galaxy's kinetic energy budget. To
account for this, we instead use the { luminosity-weighted mean}
velocity dispersion
\begin{equation}
\sigma = \frac{\int_0^{a_{\rm max}} \sigma_{\rm maj}(a) I(a)
a\,da}{\int_0^{a_{\rm max}} I(a) a\,da}, \label{sigmeq2}
\end{equation}
with $\sigma_{\rm maj}(a)$ and $I(a)$ the major-axis velocity
dispersion and the surface brightness, respectively, at a major-axis
distance $a$ and $a_{\rm max}$ the major-axis distance of the last
kinematical data-point. { Since the velocity dispersion can be
strongly influenced by the nucleus and hence show a pronounced central
depression, the luminosity-weighted mean velocity dispersion can
differ from the central dispersion by as much as 100\% of the central
value. However, for $a_{\rm max}$ larger than 1~$R_{\rm e}$, $\sigma$
changes very little by integrating farther out so we are sure that we
have defined a robust quantity.} For the other galaxies, we had to
settle for the mean (G03) and the median (vZ04) of the dispersion
profile, measured along the major axis.

As is obvious from the panel {\bf a} in Fig. \ref{LBsigma}, the FJR
becomes noticeably flatter below $\log(L_B) \sim 9.5$ or $M_B \sim
-18.3$~mag (here and in the following, $L_B$ is expressed using the
solar B-band luminosity as unit and $\sigma$ is expressed in units of
km~s$^{-1}$). We fitted a straight line to the available data, taking
into account the errors on the luminosities and the velocity
dispersions. For the luminosities, we used $\delta_{\log
L_B}=0.1$. The velocity-dispersion errors $\delta_{\log \sigma}$ of
our data can be found in Table \ref{tab1}. For the other data sets, we
of course used the errors stated by the various authors. We minimized
the non-linear quantity
\begin{equation}
\chi^2 = \frac{1}{N-2} \sum_{i=1}^N  \frac{\left(a + b\log \sigma_i - \log
{L_B}_i\right)^2}{b^2\,\delta_{\log\sigma_i}^2 + \delta_{\log {L_B}_i}^2} ,
\end{equation}
with $\log \sigma_i$ and $\log {L_B}_i$ observed data-points and
$\delta_{\log \sigma_i}$ and $\delta_{\log {L_B}_i}$ the corresponding
errorbars, using the routine {\tt fitexy} of \cite{pr92} in order to
obtain the zeropoint $a$ and the slope $b$. $N$ is the number of data
points. The diagonal elements of the estimated covariance matrix were
used as approximations to the variances of the regression coefficients
$a$ and $b$. A straight-line fit to the DR04, vZ04, G03, and B92 dEs
yields {
\begin{equation}
\log L_B = 6.02^{\pm 0.31} + 1.57^{\pm 0.19} \log \sigma
\end{equation}
with a regression coefficient $r=0.68$ and $\chi^2=4.9$. If the B92
dSphs are included in the fit, one obtains
\begin{equation}
\log L_B = 4.39^{\pm 0.37} + 2.55^{\pm 0.22} \log \sigma
\end{equation}
with a regression coefficient $r=0.85$ and $\chi^2=9.2$.} The linear
trend becomes more significant since a much larger data-interval is
covered but the $\chi^2$ is much higher, quantifying what the eye sees
immediately:~the $\sigma-L_{\rm B}$ relation does not have a constant
slope when going from the dEs to the dSphs. A dE FJR that is steeper
than the FJR of bright ellipticals is definitely excluded by these
data. This clearly shows the need of combining the available data-sets
of dE kinematics and the dangers involved with extrapolating each of
these data-sets separately towards the dSphs. Moreover, these results
contradict the universality of the $\sigma-L_{\rm B}$ relation, which
was suggested by \cite{gu93}. { A currently running project on the
properties of faint early-type galaxies ($-22 < M_R < -17.5$) in the
central 1$\deg$ of the Coma cluster yields FJR $L \propto
\sigma^{2.01}$ at the faint end, which is consistent with our results
(Matkovi\'c~\&~Guzm\'an, private communication).}

According to the YA87 models, if a mass-fraction $f$ is blown away, a
galaxy evolves to a new virial equilibrium with a velocity dispersion
that is offset from the initial value by
\begin{equation}
\Delta \log \sigma = \log(1-f) \label{dsigma1}
\end{equation}
for slow, and
\begin{equation}
\Delta \log \sigma = 0.5 \log(1-2f), \label{dsigma2}
\end{equation}
for instantaneous mass-loss.

Hence, the expansion following the galactic wind drives more {
low-mass} dEs, that lose a larger mass-fraction than more massive
objects, progressively towards lower velocity dispersions, leading to
a flatter FJR in the luminosity regime of the dEs (cf. { the shift
with respect to the zero mass-loss curve of the evolutionary endpoints
of a typical $m_B=16$ Fornax dE}, indicated by white circles in
Fig. \ref{LBsigma}). Galaxies with luminosities in the range $\log L_B
\approx 4.4-8.2$ ($M_B=-5.5$ to $-15.0$~mag)are disrupted by the
instantaneous ejection of more than half of their initial mass which
causes the velocity dispersion to drop to zero at $\log L_B \approx
8.2$ ($M_B=-15.0$~mag). The locus of the dEs in a $\log L_B$ versus
$\log \sigma$ diagram is bracketed nicely by the lines corresponding
to slow and instantaneous mass-loss. The zero mass-loss YA87
$\sigma-L_{\rm B}$ relation has a slope similar to that of the bright
ellipticals although it underestimates the velocity dispersion of the
very brightest ellipticals { (most likely due to the fact that
mergers are not taken into account)}. The CC02 models A collapse to
rather centrally concentrated systems with velocity dispersions that
are about a factor two too high. The CC02 models B agree with the YA87
models for slow mass-loss and also reproduce the observations fairly
well, from the dSphs up to the brightest ellipticals. { For a
comparison of the FJR with SAM-predictions, with and without dynamical
response to mass loss, see Figs. 12 and 14, respectively, of NY04.}

\subsection{The $R_{\rm e}-L_{\rm B}$ relation or Fish's law} \label{rlsub}

According to the YA87 models, galaxies experiencing mass-loss are
expected to evolve towards larger half-light radii at fixed
luminosity, i.e. to the right in panel {\bf b} of Fig. \ref{LBsigma},
which relates the half-light radius (in kpc) to luminosity, by an
amount
\begin{equation}
\Delta \log R_{\rm e} = -\log(1-f) \label{dre1}
\end{equation}
for slow, and
\begin{equation}
\Delta \log R_{\rm e} = \log(1-f) - \log(1-2f),\label{dre2}
\end{equation}
for instantaneous loss of a mass-fraction $f$. Galaxies in the
luminosity range $\log L_B \approx 4.4-8.2$ ($M_B \approx -5.5$ to
$-15.5$~mag) are disrupted by the instantaneous loss of more than half
of their initial mass. This causes their half-light radius to diverge
at $\log L_B \approx 8.2$ ($M_B \approx -15$~mag). The zero mass-loss
$L_{\rm B}-R_{\rm e}$ relation predicted by YA87 (i.e. the sequence
that galaxies would trace if the dynamical response to the mass loss
is not taken into account) is slightly steeper than that of the bright
elliptical galaxies and the very brightest elliptical galaxies have
larger half-light radii than is expected. The slope of the zero
mass-loss relation in combination with the dynamical response to the
galactic wind result in a dE $L_{\rm B}-$R$_{\rm e}$ relation that is
much steeper than that of the brighter galaxy species, as is actually
observed. For bulges, intermediate-luminosity and bright ellipticals,
the best fitting straight line is given by $L_B \propto R_{\rm
e}^{1.19}$ (\cite{gu93}).

For the dEs, we used the same non-linear least-squares technique as in
the previous section, assuming an error of 5\% on the { DR04
effective radii. This error takes into account the error on the
measurement of the total magnitude and photon-shot noise. It should be
noted that not all authors measure effective radii the same way. In
DR04, $R_{\rm e}$ is the radius $R_{\rm e,\,circ}$ of the circular
aperture that encloses half the light. Ba03 and Gr03 use the geometric
mean radius $R_{\rm e,\,mean}$ (which is not expected to deviate much
from $R_{\rm e,\,circ}$). G03 uses the major-axis distance $R_{\rm
e,\,maj}$ as radius. We derived geometric mean half-light radii from
the G03 data using the relation $R_{\rm e,\,mean} = \sqrt{1-\epsilon}
R_{\rm e,\,maj}$, with $\epsilon$ the ellipticity. The B92 are a mix
of $R_{\rm e,\,circ}$ and $R_{\rm e,\,mean}$ measurements.} A fit to
the DR04, G03, and B92 dEs yields {
\begin{equation}
\log L_B = 8.72^{\pm 0.04} + 2.07^{\pm 0.26} \log R_{\rm e}
\end{equation}
with a regression coefficient $r=0.68$ and $\chi^2=5.1$. Taking the
dSphs together with these dEs makes the slope even steeper:
\begin{equation}
\log L_B = 8.50^{\pm 0.05} + 3.71^{\pm 0.20} \log R_{\rm e}
\end{equation}
with a regression coefficient $r=0.90$ and $\chi^2=8.2$.} One should be
very cautious about these results since adding the Ba03 dEs clearly
destroys the impression of a strong linear relation. Also, the YA87
models do not predict that a tight $L_{\rm B}-R_{\rm e}$ relation
should exist, given the spread between the models with slow and
instantaneous mass-ejection (unlike in the case of the $\sigma-L_{\rm
B}$ relation where slow and instantaneous mass-loss models almost
coincide in the dE-regime). The slope of the dE/dSph $L_{\rm
B}-$R$_{\rm e}$ relation is reproduced very nicely by the CC02 models,
although the models of series A, that virialized at high redshift, are
too compact while the models of series B, that virialized at low
redshift, are too extended. This however does not necessarily
invalidate these calculations nor the wind model since the initial
density of the models, and hence the present-day half-light radius to
which they collapse, depends on the virialization redshift which was
arbitrarily fixed at $z=5$ by \cite{cc02}. These authors argue that,
interpolating between the CC02 models A and B, models that virialized
at $z \sim 2$ would be able to reproduce the position of dEs in this
diagram.

The locus of the observed dEs is bracketed by the YA87 models for slow
and instantaneous gas removal. The K01 models { roughly agree with
the observations and with the YA87 models. SAMs (like NY04) predict
some cosmic scatter on structural scaling relations because of the
different merger trees that lead up to a galaxy with a given
present-day luminosity. K01 only calculated one model per mass per
feedback efficiency. It is therefore not possible to assess the
uncertainty on (or the scatter about) the predicted model properties.
Still, it is encouraging that the models with minimal feedback
efficiency approximate the zero or slow mass-loss models of YA87 while
the models with high feedback efficiency expand towards larger
half-light radii, reproducing the onset of the dE sequence. For a
comparison of the $L_{\rm B}-R_{\rm e}$ relation with SAM-predictions
see Fig. 10 of NY04.}

\subsection{The $I_{\rm e}-L_{\rm B}$ relation}

The YA87 models predict that galaxies should expand after the
galactic-wind phase and evolve towards a more diffuse state,
characterised by a much lower surface brigthness. Hence, they should
move to the left in panel {\bf c} of Fig. \ref{LBsigma}, which relates
surface brightness to total luminosity, by an amount
\begin{equation}
\Delta \log I_{\rm e} = 2\log(1-f)\label{iere11}
\end{equation}
for slow, and
\begin{equation}
\Delta \log I_{\rm e} = 2\log(1-2f)-2\log(1-f),\label{iere22}
\end{equation}
for instantaneous mass-loss. Galaxies in the luminosity range $\log
L_B \approx 4.4-8.2$ ($M_B \approx -5.5$ to $-15.5$~mag) become
unbound after the instantaneous loss of more than half of their
initial mass and consequently evolve towards zero surface
brightness. In the range $\log L_B = 7-11$ ($M_B = -12$ to $-22$~mag),
the zero mass-loss $I_{\rm e}-L_{\rm B}$ relation runs almost
vertically in this diagram, predicting that $\log I_e \approx 2.5$ for
all elliptical galaxies. While most intermediate-luminosity
ellipticals indeed scatter around this value, bright elliptical
galaxies tend to be more diffuse, with surface brightnesses in the
range $\log I_e \approx 2.0-2.5$. Since less massive galaxies are
expected to lose a larger mass-fraction than more massive objects,
they move progressively towards lower surface brightnesses, changing
the slope of the $I_{\rm e}-L_{\rm B}$ relation. The YA87 models for
slow and instantaneous mass-loss are able to reproduce the observed
trend in the $I_{\rm e}-L_{\rm B}$ diagram, with the dE-sequence
running almost perpendicular to that of the bright galaxies. The
faintest K01 models also coincide with the observed locus of the
dEs. The CC02, being either too diffuse or too compact depending on
the virialization redshift, end up having too high or too low surface
brightness (see the discussion in subsection \ref{rlsub}). However,
they roughly reproduce the slope of the observed dE sequence. { For
a comparison of the $L_{\rm B}-I_{\rm e}$ relation with
SAM-predictions see Fig. 9 of NY04.}

\subsection{The $R_{\rm e}-I_{\rm e}$ or Kormendy relation}

The 12 dEs in the B92 sample suggest a $R_{\rm e}$-$I_{\rm e}$
relation that runs almost perpendicular to that of the bright
ellipticals and bulges (\cite{gu93}). However, this picture changes
completely when { more data, especially of faint galaxies, are
added, as is obvious from panel {\bf d} in Fig. \ref{LBsigma}. This
result was already derived by e.g. \cite{ca92} and \cite{gr03}.}
Mass-loss moves galaxies towards larger half-light radii and towards
much lower surface brightnesses, i.e. slightly to the right and
steeply downward with respect to the initial $\log R_{\rm e}$-$\log
I_{\rm e}$ relation, by an amount given by equations (\ref{dre1}) and
(\ref{iere11}) for slow, and equations (\ref{dre2}) and (\ref{iere22})
for instantaneous loss of a mass-fraction $f$. The region occupied by
models that are disrupted by instantaneous gas ejection is bounded by
the leftmost and rightmost thick grey lines in panel {\bf d} of
Fig. \ref{LBsigma}. The position of a typical $m_B = 16$ Fornax dE
before and after slow or fast gas removal is indicated by white
circles in Fig. \ref{LBsigma}, showing the direction in which dEs are
expected to evolve in this diagram. The slope of the dE $R_{\rm
e}$-$I_{\rm e}$ relation (if it can be called such, given the large
scatter) is nicely reproduced by these models. The $ R_{\rm
e}$-$I_{\rm e}$ relation predicted by YA87 becomes much flatter in the
regime of the intermediate-luminosity and bright elliptical
galaxies. However, the very brightest elliptical galaxies have larger
half-light radii and hence lower surface brightnesses than accounted
for by the YA87 relation, { due to not taking into account the fact
that mergers played an important role in shaping bright
ellipticals}. Again, the CC02 models are either too compact or too
extended. They roughly reproduce the trend going from dEs towards
dSphs. The photometric properties of the K01 models agree rather well
with the observations. The effect of the kinetic feedback efficiency
of supernova explosions is clear: models with a low feedback
efficiency (top sequence of K01 models in panel {\bf d} of
Fig. \ref{LBsigma}) experience a late galactic wind and lose little
gas. Consequently, the dynamic response is rather mild { and they
scatter about the YA87 zero mass-loss curve}. Models with a high
feedback efficiency (bottom sequence in panel {\bf d} of
Fig. \ref{LBsigma}) lose much more gas and expand towards much larger
$R_{\rm e}$ and lower $I_{\rm e}$. At fixed mass but for different
feedback efficiencies, the trend is in the same direction as the
observed $R_{\rm e}$-$I_{\rm e}$ relation and as the YA87 models (with
feedback efficiency replaced with mass-loss timescale). { For a
comparison of the Kormendy relation with SAM-predictions see Fig. 11
of NY04.}




\section{Bivariate relations} \label{bivar}

\subsection{The Fundamental Plane (FP)}

\begin{figure*}
\vspace*{9cm}
\special{hscale=55 vscale=55 hsize=550 vsize=550
hoffset=-10 voffset=-150 angle=0 psfile="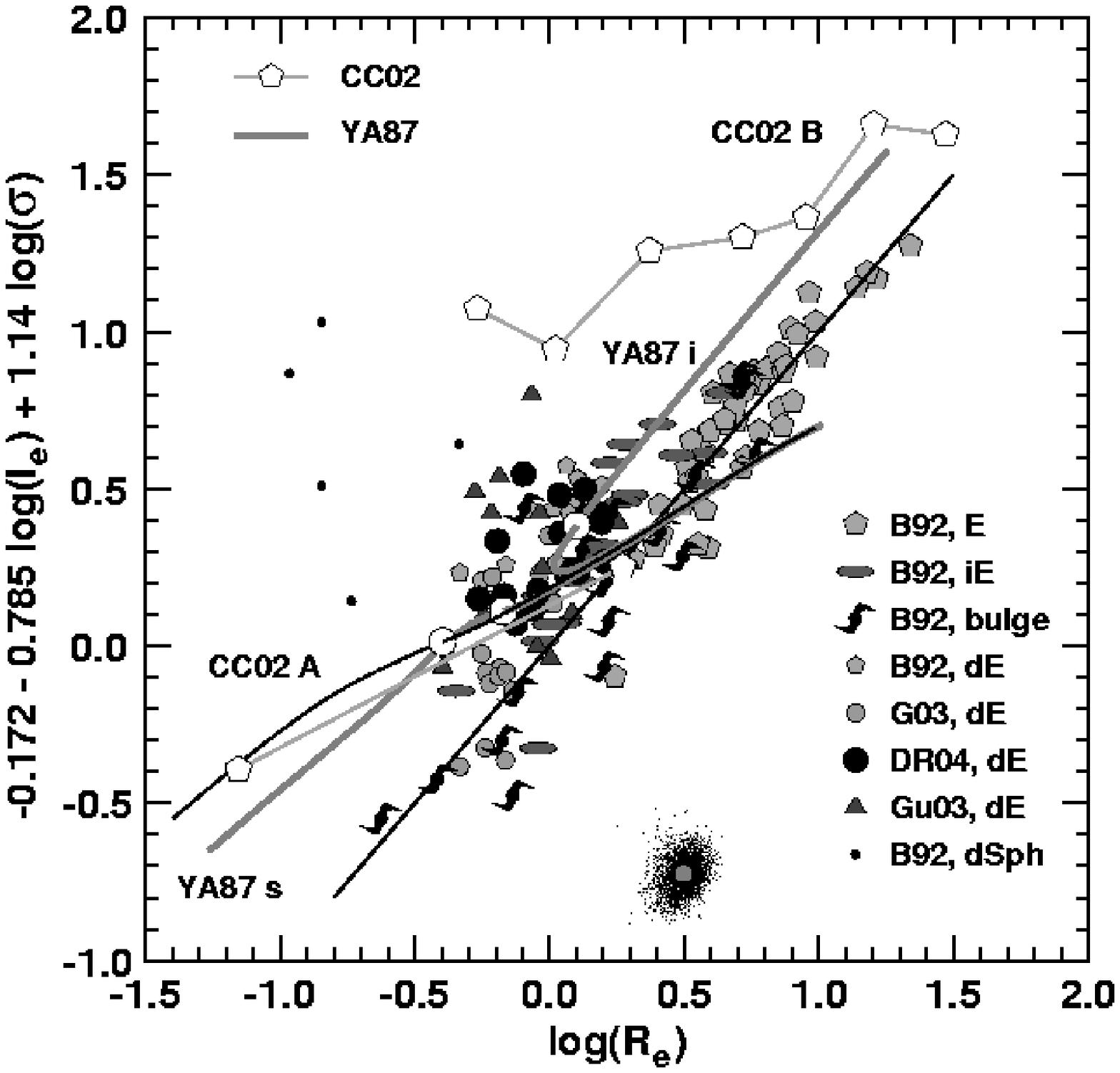"}
\special{hscale=55 vscale=55 hsize=550 vsize=550
hoffset=250 voffset=-150 angle=0 psfile="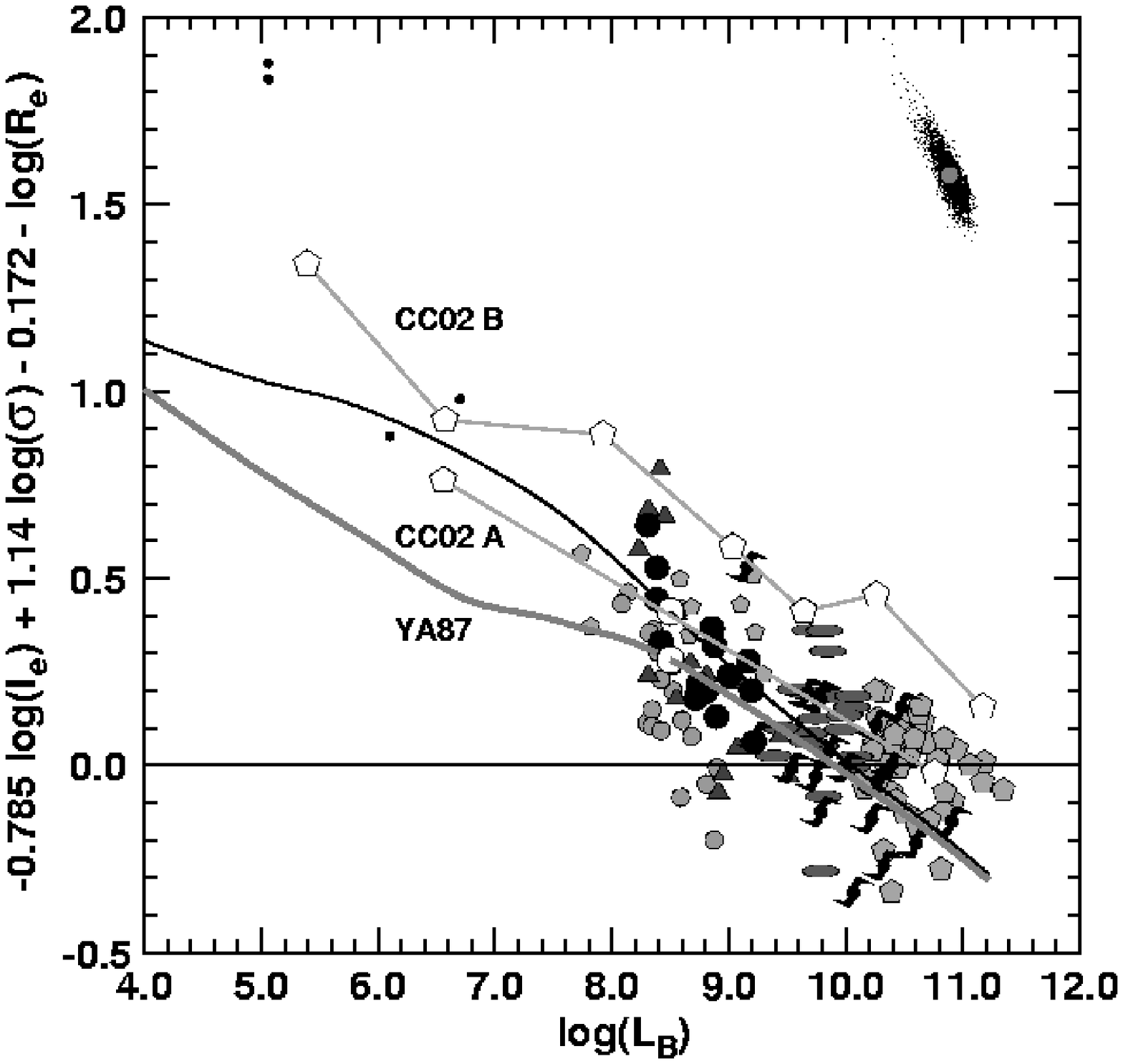"}
\caption{Left panel:~an edge-on view onto the FP of the bulges, bright
and intermediate-luminosity ellipticals in ($\log R_{\rm e},\log
I_{\rm e},\log \sigma$) space. All symbols have the same meaning as in
Fig. \ref{LBsigma}. The straight black line traces the $\log R_{\rm e}
= -0.172 - 0.785 \log I_{\rm e}+ 1.14 \log \sigma$ relation that
defines the FP. The dEs observed by \cite{gu03} (Gu03) have been added
to the DR04, G03, and B92 data sets. The curved black line traces the
FP of the zero mass-loss YA87 models. The thick grey lines give the
locus of the YA87 models afer slow (bottom thick grey line) or fast
(top thick grey line) gas removal. The white circles show the expected
evolution of a $m_B=16$ Fornax dE:~most of the evolution happens
almost parallel to the FP so that, given the slope of the zero
mass-loss YA87 FP, one would expect dEs to lie slightly above the FP
defined by the bright ellipticals { in this projection}. Right
panel:~vertical deviation of the FP as a function of luminosity. All
models predict less massive galaxies to lie progressively higher above
the FP { in this projection}, as is observed. { The YA87 curves
for slow and instantaneous winds coincide in this diagram.} In each
panel, a cloud of 3000 simulated { data points} is plotted to give
an idea of the systematic and statistical uncertainty on a typical
{ DR04} data point (indicated by a grey dot).
\label{FP1} }
\end{figure*}

The FP of the bulges and the bright and intermediate-luminosity
ellipticals is given by the equation
\begin{equation}
\log R_{\rm e} = {\rm const. } + 1.14 \log \sigma - 0.785 \log I_{\rm
e},
\end{equation}
(\cite{gu93}). The FP can be seen as an emanation of the virial
theorem, relating the potential energy $\cal{PE}$ and kinetic energy
$\cal{KE}$ budgets of a galaxy in equilibrium. More specifically, with
${\cal PE} = GM^2/R_G$ and ${ \cal KE} = \frac{1}{2} M \langle v^2
\rangle$, where $M$ is the mass of the galaxy, $R_G$ its gravitational
radius, and $\langle v^2 \rangle$ the mass-weighted mean-square
stellar velocity (see \cite{bt87}), one finds that ${\cal PE}+2 {\cal
KE}=0$ or that $M \propto R_G \langle v^2 \rangle$. Hence, one expects
that $M = c \,R_{\rm e} \sigma^2$, with $c$ a proportionality
parameter that (hopefully) transforms the mass-weighted theoretical
parameters ($R_G$, $\sqrt{\langle v^2 \rangle}$) to the {
luminosity-weighted} observed ones ($R_{\rm e}$, $\sigma$). This,
together with the relation $L_B = 2 \pi I_{\rm e} R^2_{\rm e}$, leads
to the theoretical FP relation
\begin{equation}
\log R_{\rm e} = {\rm const.} + \log c - \log
\frac{M}{L} + 2 \log \sigma - \log I_{\rm e}.
\end{equation}
Variations of $c$ (structural non-homology) and/or the mass-to-light
ratio $M/L$ along the mass sequence of dynamically hot galaxies,
exacerbated by an imperfect correspondence between $\sigma$ and
$\sqrt{\langle v^2 \rangle}$ due to rotation and anisotropy, slightly
change the tilt of the FP to the observed value and introduce some
scatter (see e.g. \cite{tr04}).

The positions of the various galaxy species in a side view of the FP
are plotted in Fig. \ref{FP1}. We did not correct our velocity
dispersions for aperture effects in this plot since applying the
prescriptions given by \cite{bl02} leads to corrections of order
$\Delta \log\sigma = 0.02$ and are clearly negligibly small. In both
panels of Fig. \ref{FP1}, a cloud of 3000 simulated data points,
calculated along the lines discussed in section \ref{mono}, is plotted
to give an idea of the systematic and statistical uncertainty on a
typical { DR04} data point (indicated by a grey dot). Here and in
the following, we have also made use of the FP-data of 14 Coma cluster
dEs presented by \cite{gu03} (Gu03). By and large, dEs show a tendency
to lie above the FP { in this projection}. The Local Group dSphs
deviate even more extremely from the FP. The dynamical response after
the loss of a mass-fraction $f$, given by
\begin{eqnarray}
\Delta \left( - 0.785 \log I_{\rm e} + 1.14 \log \sigma \right) &=&
-0.43 \log (1-f), \nonumber \\ \Delta \log R_{\rm e} &=& -\log(1-f),
\end{eqnarray}
for slow removal and
\begin{eqnarray}
\Delta \left( - 0.785 \log I_{\rm e} + 1.14 \log \sigma \right) &=&
1.57 \log(1-f) \nonumber \\ && \,\,\,\,\,\,\,\,\,\,\,\,\,\,\,\,\,\,\,-
\log (1-2f), \nonumber \\ \Delta \log R_{\rm e} &=& \log(1-f) -
\log(1-2f),
\end{eqnarray}
for instantaneous mass-loss, moves dEs almost parallel to the FP in
the direction of larger half-light radii (see e.g. the typical
$m_B=16$ Fornax dE which moves from left to right, as shown by the
white circles in Fig. \ref{FP1}). Since the zero mass-loss FP has a
much flatter slope than the observed one, fainter galaxies are indeed
expected to lie progressively above the FP { in this
projection}. The CC02 models A agree very well with the YA87 models
and with the observations. The surface brightness of the CC02 models B
however is much lower than observed, placing them well above the
observed FP { in this projection}.

In the right panel of Fig. \ref{FP1}, the deviations of the different
galaxy species from the FP are plotted as a function of
luminosity. The expected change of the FP residual after the ejection
of a mass fraction $f$ is given by
\begin{eqnarray}
\Delta \left( (-0.785 \log I_{\rm e} + 1.14 \log \sigma ) - \log
R_{\rm e}\right) &=& \nonumber \\ && \!\!\!\!\!\!\!\!\!\!\!\!0.57 \log(1-f),
\end{eqnarray}
i.e. slightly downwards with respect to the YA87 zero mass-loss
sequence, which itself is a steep function of luminosity. This
expression is valid both for slow and instantaneous winds although in
the latter case galaxies in the luminosity range $\log L_B \approx
4.4-8.2$ ($M_B \approx -5.5$ to $-15.5$~mag) cease to exist. Hence,
according to the YA87 calculations, the dynamical response after
mass-loss in fact reduces the FP residual. Still, dEs are expected to
lie above the FP { in this projection}. The YA87 and the CC02
models A roughly reproduce the observed trend as a function of $\log
L_B$. The CC02 models of series B show approximately the same trend
but lie systematically above the FP { in this projection}.

\subsection{The FP in $\kappa$-space}
\begin{figure*}
\vspace*{9cm}
\special{hscale=52 vscale=52 hsize=550 vsize=550
hoffset=-46 voffset=-150 angle=0 psfile="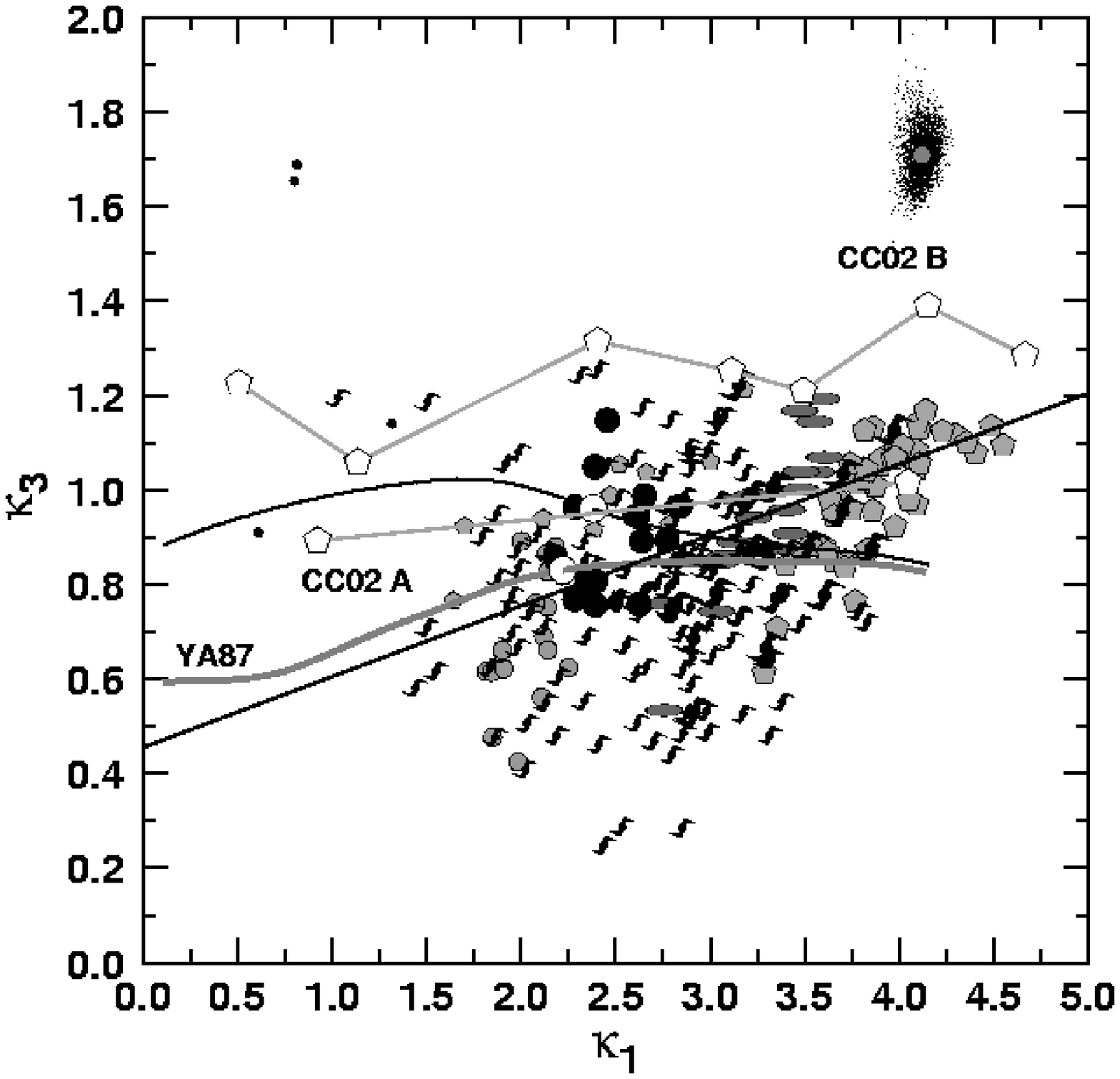"}
\special{hscale=52 vscale=52 hsize=550 vsize=550
hoffset=227 voffset=-150 angle=0 psfile="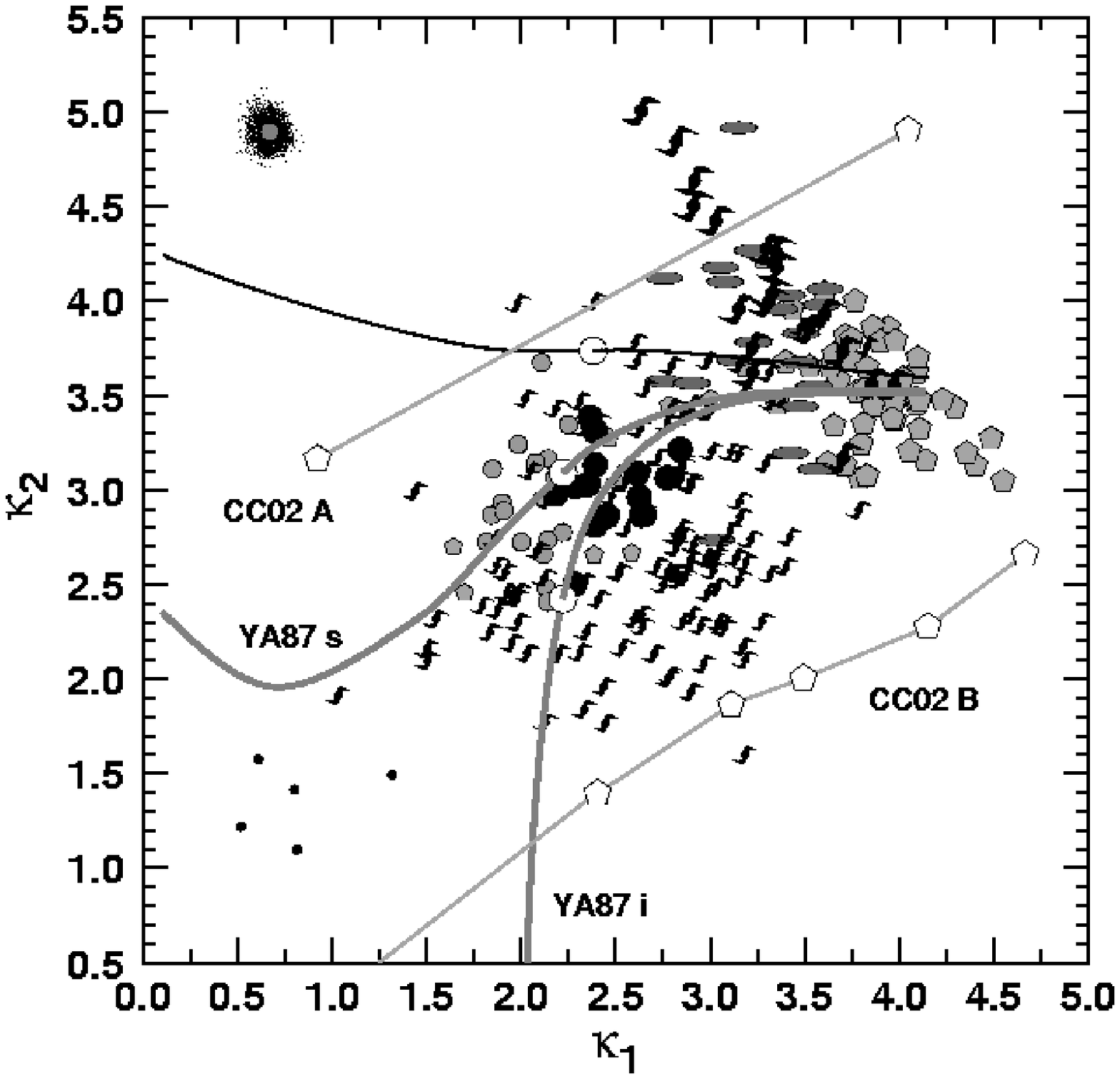"}
\caption{The FP in $\kappa$-space. Left panel:~edge-on view of the FP,
the black straight line traces the $\kappa_3 = 0.46 + 0.15 \kappa_1$
relation. All symbols have the same meaning as in
Fig. \ref{LBsigma}. The Sc/Sd/Irr galaxies taken from \cite{bu97} are
plotted as small spirals. According to the YA87 models, the evolution
in the ($\kappa_1,\kappa_3$) plane does not depend on the time-scale
on which mass is lost as a galactic wind. The predicted tendency for
faint dwarf galaxies to lie above the FP { in this projection} is
borne out by the observations. Right panel:~face-on view onto the FP
{ (equivalent to panel {\bf c} of Fig. \ref{LBsigma})}. The
evolution within the FP is much more outspoken:~the mass-loss suffered
by the dwarf galaxies makes them expand, shifting them towards lower
$\kappa_2$ values. The dEs and dSphs end up forming a sequence almost
parallel to that of the bulges and the bright and
intermediate-luminosity ellipticals. The CC02 models A reproduce the
slope of the dE FP but are somewhat too compact and as a consequence
have too high $\kappa_2$ values. The CC02 models B on the contrary are
too extended lie below the dE sequence in the ($\kappa_1,\kappa_2$)
plane. Still, all models predict the same slope of the dE sequence in
the ($\kappa_1,\kappa_2$) plane. In each panel, a cloud of 3000
simulated { data points} is plotted to give an idea of the
systematic and statistical uncertainty on a typical { DR04} data
point (indicated by a grey dot). \label{kappa}}
\end{figure*}

Another way to view the FP is in the so-called $\kappa$-space
(\cite{be92}) defined by
\begin{eqnarray} 
\kappa_1 &=& \frac{1}{\sqrt{2}} \log\left(R_{\rm e} \sigma^2\right),
\,\, \kappa_2 = \frac{1}{\sqrt{6}} \log\left(\frac{I^2_{\rm
e} \sigma^2}{R_{\rm e}} \right), \nonumber \\ \kappa_3 &=&
\frac{1}{\sqrt{3}} \log\left( \frac{\sigma^2}{I_{\rm e} R_{\rm e}}
\right).
\end{eqnarray}
Using these definitions and the virial theorem, $\kappa_1$ is expected
to be a measure for the mass of a galaxy, $\kappa_2$ is sensitive
mostly to surface brightness and $\kappa_3$ depends on the
mass-to-light ratio. A galaxy is shifted in $\kappa$-space by an
amount
\begin{eqnarray}
\Delta \kappa_1 &=& \frac{1}{\sqrt{2}} \log (1-f), \, \,
\Delta \kappa_2 = \frac{7}{\sqrt{6}} \log (1-f), \nonumber \\
\Delta \kappa_3 &=& \frac{1}{\sqrt{3}} \log (1-f)
\end{eqnarray}
if a mass-fraction $f$ is blown away in a slow wind. An instantaneous
wind engenders a dynamical response given by 
\begin{eqnarray}
\Delta \kappa_1 &=& \frac{1}{\sqrt{2}} \log (1-f), \nonumber \\ 
\Delta \kappa_2 &=& \frac{1}{\sqrt{6}} \left( 6\log (1-2f)-5\log (1-f) \right), \nonumber \\
\Delta \kappa_3 &=& \frac{1}{\sqrt{3}} \log (1-f).
\end{eqnarray}
Since a galaxy's luminosity is not affected by the ejection of gas,
$I_{\rm e} \propto R_{\rm e}^{-2}$, and
\begin{equation}
\Delta \kappa_1 \propto \Delta \kappa_3 \propto \Delta \log M.
\end{equation}
Galaxies are therefore expected to evolve downwards and to the left in
the ($\kappa_1,\kappa_3$)-projection of the FP (left panel of
Fig. \ref{kappa}).  In both panels of Fig. \ref{kappa}, a cloud of
3000 simulated { data points}, calculated along the lines discussed
in section \ref{mono}, is plotted to give an idea of the systematic
and statistical uncertainty on a typical { DR04} data point
(indicated by a grey dot). The YA87 zero mass-loss FP however has a
much flatter slope than the observed FP, which tends to position dEs
slightly above the FP { in this projection}. As in the right panel
of Fig. \ref{FP1}, the dynamical response to the galactic wind moves
galaxies away from the zero mass-loss FP towards the FP of the bright
galaxies.

The sensitivity of $\kappa_2$ to surface brightness, which drops
significantly as a galaxy expands after the wind phase, leads to a
strong downward evolution in the ($\kappa_1,\kappa_2$)-projection of
the FP (right panel of Fig. \ref{kappa}). The models undergoing an
instantaneous wind are disheveled most severely by the suffered
mass-loss and show the most pronounced evolution in the
($\kappa_1,\kappa_2$) plane. The YA87 models and the CC02 models A are
able to reproduce the position of the dEs in the
($\kappa_1,\kappa_3$)-projection of the FP. The CC02 models B are too
extended and consequently lie above the observed FP { in this
projection}. In the ($\kappa_1,\kappa_2$)-projection, which is a
nearly face-on view onto the FP, the dEs define a sequence running
almost perpendicular to that of the bulges and the bright and
intermediate-luminosity ellipticals. Within the context of the YA87
models, this is in part due to the zero mass-loss sequence which runs
almost horizontally across the ($\kappa_1,\kappa_2$)-plane and to the
evolution induced by the mass-loss during the galactic-wind phase,
which shifts galaxies downwards. Both the YA87 and the CC02 models
reproduce the observed slope of the dE sequence in the
($\kappa_1,\kappa_2$)-plane.

The present-day Sc/Sd/Irr galaxies are probably the stellar systems
that best resemble the dE progenitors as envisaged by the harassment
scenario. Their position in $\kappa$-space, with the maximum rotation
velocity $v_{\rm max}$ used as a substitute for the theoretical
quantity $\sqrt{\langle v^2 \rangle}$ in the virial theorem, is
indicated in Fig. \ref{kappa} by small spiral symbols. This population
of late-type galaxies has a much larger scatter in $\kappa$-space than
the dEs. It is not clear how harassment, which is by nature a
stochastic process whose effects depend in a very complicated way on
the orbit of a progenitor galaxy through a given galaxy cluster or
group, could transform the scatter cloud of late-types into the much
tighter dE sequence.


\section{Discussion and conclusions} \label{discuss}

We have presented the sequences traced by dEs in the $\log L_B$
vs. $\log \sigma$, $\log L_B$ vs. $\log R_{\rm e}$, $\log L_B$
vs. $\log I_{\rm e}$, and $\log R_{\rm e}$ vs. $\log I_{\rm e}$
diagrams and in the ($\log \sigma,\log R_{\rm e},\log I_{\rm e}$)
parameter space in which bright and intermediate-luminosity elliptical
galaxies and bulges of spirals define a Fundamental Plane (FP). These
results are based on three equally large kinematical data-sets:~the
data presented in this paper (DR04), G03, and vZ04. This brings the
number of dEs with resolved kinematics to 43. More data were added
when studying correlations involving only photometric data (Ba03 {
and Gr03}) or the FP (Gu03). { We also used the kinematical and
photometric data presented in B92.} Our main conclusions are the
following:
\begin{itemize}
\item We have shown that the $\sigma-L_{\rm B}$ or Faber-Jackson
relation does {\em not} have a constant slope when going from the
bright ellipticals, over the dEs, down to the dSphs, contrary to
previous claims. The $L_B \propto \sigma^{3.7}$ relation of the bright
galaxies changes into a $L_B \propto \sigma^{1.6 \pm 0.2}$ relation
below $L_B \approx 9.4$ ($M_B \approx -18$~mag). A dE FJR as steep as
or even steeper than that of the bright ellipticals is definitely
excluded by these data. A flatter slope is predicted by all models
presented here (YA87, CC02) as a consequence of the dynamical response
to the supernova-driven gas ejection. More sophisticated SAMs that
simulate the hierarchical merger tree that leads up to the formation
of a galaxy show that post-merger starbursts are absolutely necessary
to bring the models in agreement with a whole host of observations and
that a significant flattening of the FJR in the dE-regime is a direct
consequence of the dynamical response to starburst-induced gas
ejections (\cite{so01}, \cite{ny04}). As a consequence, dEs and their
progenitors, being low-mass, fragile objects, are the ideal objects to
study if one aims at further refining prescriptions for
star-formation, supernova feedback and the response to galactic winds
in semi-analytical or numerical simulations of cosmological structure
growth. { It should be noted however that gas-rich dwarf late-type
galaxies exist whose large H{\sc i} content appears to be at variance
with CDM predictions (\cite{mm04}). It is not clear yet how worrisome
this actually is and, since we restrict ourselves to the properties of
gas-poor dEs, a discussion of the H{\sc i} properties of dwarf
late-type galaxies is clearly beyond the scope of this paper.}
\item The simple fact that these diffuse, low surface-brightness, low
velocity-dispersion dEs exist puts strong constraints on the redshift
dependence of the cosmic star-formation rate. \cite{ny04} have shown
convincingly that structure-formation models in a $\Lambda$CDM
universe with a short star-formation timescale at high $z$ fail to
produce such inflated dEs. In such a universe, dEs are assembled from
progenitors that have already converted most of their gas into stars,
and they are expected to trace the same sequences in the panels of
Fig. \ref{LBsigma} as the giant ellipticals (which are formed further
down the merger tree from almost purely stellar progenitors,
independent of the cosmic star-formation rate). Only models that have
long enough star-formation timescales at high $z$, such that dEs can
be formed by the mergers of gaseous progenitors, agree with these
observations { (e.g. the models of NY04 with a redshift independent
star-formation time scale $\tau_*=1.3$~Gyr, where the star-formation
rate is given by $M_{\rm gas}/{\tau_*}$ with $M_{\rm gas}$ the H{\sc
i} mass)}. The starburst triggered by each merger and the ensuing
supernova-explosions eject gas and thus lead to a population of
diffuse dwarf galaxies with low velocity dispersions, as observed.
{ This also explains the success of the YA87 and CC02 models, which
do not take into account mergers but rather assume galaxies to
originate from a single gas cloud, for dEs while they fail to
reproduce the properties of massive ellipticals.}

However, all this needs to be reconciled with the observation that the
stellar mass density of the universe at redshift $z>2$, contained in
massive galaxies, is larger than can be accounted for by SAMs
(\cite{fo03}, \cite{so04}). Still, the distribution of the dEs in the
various univariate diagrams can, if enough dwarf galaxies have been
observed for statistically sound statements to be made, put very
stringent limits on the redshift dependence of the star-formation
timescale in semi-analytical or numerical simulations of galaxy
evolution.
\item Models for the evolution of dwarf and intermediate-luminosity
elliptical galaxies, based on the idea that these stellar systems grow
from collapsing primordial density fluctuations, are able to reproduce
the observed relations between parameters that quantify their
structure ($L_B$, $R_{\rm e}$, $I_{\rm e}$) and internal dynamics
($\sigma$) quite well. Despite their simplicity, the YA87 models
account very well for the behaviour of dEs in the ($\log \sigma,\log
R_{\rm e},\log I_{\rm e}$) parameter space. Although numerically and
physically much more sophisticated and with a more sound cosmological
footing, the predictions of the models presented by K01 and
\cite{ny04} agree with the YA87 models. The CC02 models, in which the
virialization redshift of the models is set ``by hand'', fail to get
the zeropoints correctly but nonetheless are able to reproduce the
slopes of the various relations. All this suggests that little merging
took place in the life of a dE {\em after} it started forming stars. 
\item Low-mass systems such as dEs and dSphs lie above the FP { in
the projection used in Figs. \ref{FP1} and \ref{kappa}} defined by the
bright ellipticals, with fainter galaxies lying progressively higher
above the FP. It is well known that the remnant of the merger of two
galaxies is more diffuse than its progenitors (see e.g. \cite{he92}
and \cite{da03} for dissipationless mergers and \cite{he93} and
\cite{be98} for merger simulations taking into account the presence of
gas). Hence, the structural properties of the brightest elliptical
galaxies, including the slope of the FP, can be explained quite well
if late mergers (with progenitors that had already converted a
significant fraction of their gas into stars) played an important role
in their formation and evolution. This is much less the case for less
massive systems such as dEs, which formed higher up the hierarchical
merging tree from more dissipative mergers.
\item While dEs follow well-defined sequences in the various
univariate diagrams, the correlations are not as tight as in the case
of bright ellipticals. This cannot only be due to measurement
uncertainties (e.g. very deep photometry is now available that allows
to determine $L_B$ and $R_{\rm e}$ with very small errors, still the
scatter on the dE $\log R_{\rm e}$-$\log I_{\rm e}$ relation is
large). This cosmic scatter may be a consequence of the sensitivity of
these low-mass systems to both internal (supernova explosions,
feedback efficiency, the details of galactic winds, \ldots) and
external processes (gravitational interactions, tidal stripping of
stars and ram-pressure stripping of gas, \ldots) in group and cluster
environments. Hence, these objects are ideal laboratories to study
these physical processes to which bright ellipticals seem to be quite
insensitive.
\item The wind model has passed this test. However, these findings do
not necessarily falsify the harassment scenario. The dEs observed so
far overlap in $\kappa$-space (see Fig. \ref{kappa}) with the
present-day analogs of possible dE progenitors (the Scd and Irr
galaxies). This overlap leaves open the possibility that we have
observed dEs that formed via hierarchical merging {\em and} dEs that
formed via harassment. Especially since some of the dEs observed during
the Large Program provide us with very strong evidence (such as
embedded stellar disks or kinematically decoupled cores) that
harassment has indeed played an important role in their past evolution.
\end{itemize}
Hence, judging from the photometric and kinematical data that are now
available, dEs are most likely a mixed population, with primordial and
more recently (trans)formed objects co-existing in the present-day
universe. More spectroscopic age and metallicity estimates and
kinematical data are required in order to allow the importance of the
evolutionary avenues to be constrained, especially of faint, low
surface-brightness dEs in order to fill the gap between the dEs and
the dSphs in the FJR (the faintest dEs with resolved kinematics are
still $\sim 20$ times brighter than the brightest Local Group dSph)
and to determine the position of such faint dEs in the other uni- and
bivariate diagrams.

\begin{acknowledgements}
Based on observations made at the European Southern Observatory, Chile
(ESO Large Programme Nr.~165.N-0115). SDR wishes to thank Philippe
Prugniel for fruitful discussions while visiting the CRAL-Observatoire
de Lyon. WWZ acknowledges the support of the Austrian Science Fund
(project P14753). DM acknowledges the support of the Bijzonder
OnderzoeksFonds (BOF Universiteit Gent). This research has made use of
the NASA/IPAC Extragalactic Database (NED) which is operated by the
Jet Propulsion Laboratory, California Institute of Technology, under
contract with the National Aeronautics and Space Administration.
\end{acknowledgements}

\end{document}